\documentstyle[12pt]{article}

\newcommand \beq{\begin{eqnarray}}
\newcommand \eeq{\end{eqnarray}}
\begin{document}
\input epsf

%------------------------------------------------------------------------------

\def\bfgamma{\mbox{\boldmath$\gamma$}}
\def\bfalpha{\mbox{\boldmath$\alpha$}}
\def\bftau{\mbox{\boldmath$\tau$}}
\def\bfgrad{\mbox{\boldmath$\nabla$}}
\def\bfsigma{\mbox{\boldmath$\sigma$}}
\def\BN{\hbox{Bloch-Nordsiek}}
\def\vp{\mbox{$\bf v\cdot p$}}
\def\vq{\mbox{$\bf v\cdot q$}}
\def\vpq{\mbox{$\bf v\cdot(p+ q)$}}
\def\tilA{\mbox{$v\cdot A$}}
\def\tilQ{\mbox{v\cdot q}}
\def\tilQ1{\mbox{$v\cdot q_1$}}
\def\tilQ2{\mbox{$v\cdot q_2$}}
\def\bfp{\mbox{\boldmath$p$}}

% ---------- Gradient, etc.
\def\square{\hbox{{$\sqcup$}\llap{$\sqcap$}}}   % box
\def\grad{\nabla}                               % gradient
\def\del{\partial}                              % synonym for \partial

% ---------- Fractions.
\def\frac#1#2{{#1 \over #2}}
\def\smallfrac#1#2{{\scriptstyle {#1 \over #2}}}
\def\half{\ifinner {\scriptstyle {1 \over 2}}
   \else {1 \over 2} \fi}

% ---------- Bras and kets
\def\bra#1{\langle#1\vert}              % \bra{stuff} gives <stuff|
\def\ket#1{\vert#1\rangle}              % \ket{stuff} gives |stuff>

%       \simge and \simle make the "greater than about" and the "less
% than about" symbols with spacing as relations.
\def\simge{\mathrel{%
   \rlap{\raise 0.511ex \hbox{$>$}}{\lower 0.511ex \hbox{$\sim$}}}}
\def\simle{\mathrel{
   \rlap{\raise 0.511ex \hbox{$<$}}{\lower 0.511ex \hbox{$\sim$}}}}

%       \parenbar puts a bar in small parentheses over a character to
% indicate an optional antiparticle. \nunubar and \ppbar are special
% cases.

\def\parenbar#1{{\null\!                        % left-hand spacing
   \mathop#1\limits^{\hbox{\fiverm (--)}}       % (--) in 5pt type
   \!\null}}                                    % right-hand spacing
\def\nunubar{\parenbar{\nu}}
\def\ppbar{\parenbar{p}}

%       \buildchar makes a compound symbol, placing #2 above #1 and #3
% below it with \limits. \overcirc is a special case.

\def\buildchar#1#2#3{{\null\!                   % \null, cancel space
   \mathop#1\limits^{#2}_{#3}                   % #1, #2 above, #3 below
   \!\null}}                                    % cancel space, \null
\def\overcirc#1{\buildchar{#1}{\circ}{}}

%  \slashchar puts a slash through a character to represent contraction
%  with Dirac matrices. Use \not instead for negation of relations, and use
%  \hbar for hbar.

\def\slashchar#1{\setbox0=\hbox{$#1$}           % set a box for #1 
   \dimen0=\wd0                                 % and get its size
   \setbox1=\hbox{/} \dimen1=\wd1               % get size of /
   \ifdim\dimen0>\dimen1                        % #1 is bigger
      \rlap{\hbox to \dimen0{\hfil/\hfil}}      % so center / in box
      #1                                        % and print #1
   \else                                        % / is bigger
      \rlap{\hbox to \dimen1{\hfil$#1$\hfil}}   % so center #1
      /                                         % and print /
   \fi}                                         %

%       \subrightarrow#1 puts the text #1 under an arrow of the 
% appropriate length.

\def\subrightarrow#1{%                          % #1 under arrow
  \setbox0=\hbox{%                              % set a box
    $\displaystyle\mathop{}%                    % no mathop
    \limits_{#1}$}%                             % just limits
  \dimen0=\wd0%                                 % get width
  \advance \dimen0 by .5em%                     % add a bit
  \mathrel{%                                    % space like =
    \mathop{\hbox to \dimen0{\rightarrowfill}}% % arrow to width
       \limits_{#1}}}                           % text below

%----------- References
%
%\def\journal#1#2#3#4{{\sl #1}\ $\underline{\hbox{#2}}$, {#3} ({#4})}
\def\journal#1#2#3#4{\ {#1}{\bf #2} ({#3})\  {#4}}
%     Specific journals:

\def\AdvPhys{\journal{Adv.\ Phys.}}
\def\AnnPhys{\journal{Ann.\ Phys. }}
\def\EurophysLett{\journal{Europhys.\ Lett.}}
\def\JApplPhys{\journal{J.\ Appl.\ Phys.}}
\def\JMathPhys{\journal{J.\ Math.\ Phys.}}
\def\LettNuovoCimento{\journal{Lett.\ Nuovo Cimento}}
\def\Nature{\journal{Nature}}
\def\NPA{\journal{Nucl.\ Phys.\ {\bf A}}}
\def\NPB{\journal{Nucl.\ Phys.\ {\bf B}}}
\def\NuovoCimento{\journal{Nuovo Cimento}}
\def\Physica{\journal{Physica}}
\def\PLA{\journal{Phys.\ Lett.\ {\bf A}}}
\def\PLB{\journal{Phys.\ Lett.\ {\bf B}}}
\def\PhysRev{\journal{Phys.\ Rev.}}
\def\PR{\journal{Phys.\ Rev.}}
\def\PRC{\journal{Phys.\ Rev.\ {\bf C}}}
\def\PRD{\journal{Phys.\ Rev.\ {\bf D}}}
\def\PRB{\journal{Phys.\ Rev.\ {\bf B}}}
\def\PRL{\journal{Phys.\ Rev.\ Lett. }}
\def\PhysRept{\journal{Phys.\ Repts. }}
\def\ProcNatlAcadSci{\journal{Proc.\ Natl.\ Acad.\ Sci.}}
\def\ProcRoySoc{\journal{Proc.\ Roy.\ Soc.\ London Ser.\ A}}
\def\RevModPhys{\journal{Rev.\ Mod.\ Phys. }}
\def\Science{\journal{Science}}
\def\SovPhysJETP{\journal{Sov.\ Phys.\ JETP}}
\def\SovPhysJETPLett{\journal{Sov.\ Phys.\ JETP Lett.}}
\def\SovJNuclPhys{\journal{Sov.\ J.\ Nucl.\ Phys.}}
\def\SovPhysDoklady{\journal{Sov.\ Phys.\ Doklady}}
\def\ZPhys{\journal{Z.\ Phys.}}
\def\ZPhysA{\journal{Z.\ Phys.\ A}}
\def\ZPhysB{\journal{Z.\ Phys.\ B}}
\def\ZPhysC{\journal{Z.\ Phys.\ C}}

%%%%%%%%%%%% titlepage %%%%%%%%%%%%%%%%%%%%%%%%%%%%%
\begin{titlepage}
\begin{flushright}
CERN-TH/99-172\\
Saclay-T99/059 \\
hep-ph/9906485 
\end{flushright}
\vspace*{1.2cm}
\begin{center}
\baselineskip=13pt
{\Large{\bf Ultrasoft Amplitudes in Hot QCD}}
\vskip0.5cm
Jean-Paul BLAIZOT\footnote{E-mail: blaizot@spht.saclay.cea.fr}

{\it Service de Physique Th\'eorique\footnote{Laboratoire de la Direction
des
Sciences de la Mati\`ere du Commissariat \`a l'Energie
Atomique}, CE-Saclay \\ 91191 Gif-sur-Yvette, France}
\vskip0.3cm
  and 

\vskip0.3cm
Edmond IANCU\footnote{E-mail: edmond.iancu@cern.ch}

{\it Theory Division, CERN\\ CH-1211, Geneva 23, 
Switzerland}

\end{center}

\vskip 1cm
\begin{abstract} 

By using the Boltzmann equation describing the relaxation of colour
excitations in the QCD plasma, we obtain effective
amplitudes for the ultrasoft colour fields
carrying momenta of order $g^2 T$. These
amplitudes are of the same order in $g$ as the hard thermal loops
(HTL), which they generalize by including the effects of the collisions
among the hard particles. The ultrasoft amplitudes share many of the
remarkable properties of the HTL's: they are gauge invariant, obey
simple Ward identities, and, in the static limit, reduce to the usual
Debye mass for the electric fields. However, unlike the HTL's,
which correspond effectively to one-loop diagrams, the
ultrasoft amplitudes resum an infinite number of diagrams
of the bare perturbation theory.   By solving the linearized
Boltzmann equation, we obtain a formula for the colour conductivity
which accounts for the contributions of the hard and soft modes beyond the
leading logarithmic approximation.

 \end{abstract}
\vskip 1.cm

\begin{flushleft}
To appear in Nuclear Physics B
\end{flushleft}
\end{titlepage}

\setcounter{equation}{0}
\section{Introduction}

In recent years,  kinetic theory has proven to be a powerful tool to construct
effective  theories for the soft fields in ultrarelativistic plasmas.
Thus, the effective theory at the scale $gT$ follows from a
collisionless kinetic equation, of the Vlasov type
\cite{qcd}. The effective theory at the scale $g^2T$
is generated by a Boltzmann equation which includes a collision
term for colour relaxation \cite{Bodeker98,ASY98,BE,Manuel99}.
(Here, $T$ is the temperature, and $g$ is the coupling constant,
assumed to be small.) The kinetic description relies on the separation
of scales between single-particle and collective excitations.
This allows for kinematical approximations which, like the
relevant scales themselves, are controlled by powers of $g$.
By using these approximations, kinetic equations have been
rigorously constructed from the quantum equations of motion 
\cite{qcd,Bodeker98,BE,prept}, thus providing justification for
numerous previous works using ad hoc transport
equations inspired by classical physics
\cite{Baym90,Gyulassy93,Heisel94,Manuel94,Heisel94a,Baym97,Manuel99,Basag99,Markov95}.
Previous attempts to derive 
these equations \cite{EHPRept}
 generally failed to recognize the proper separation of scales  which turns out
to be essential in order to control the gauge invariance of the approximations
involved.

The single-particle excitations of the QCD plasma
are {\it hard} transverse gluons\footnote{We consider
here a purely Yang-Mills plasma, with no quarks.}
with typical momenta $k\sim T$. The {\it soft} fields
are colour fields $A^\mu_a(x)$ with momenta of the order $gT$ or less.
When acting on the hard particles, these soft fields induce
longwavelength ($\lambda \simge 1/gT$) collective excitations,
with $\lambda$ much larger than
the mean interparticle distance $\bar r \sim 1/T$
\cite{BIO96,MLB96,prept}.
In the framework of the kinetic theory, these excitations
are described by a colour density matrix $\delta N_{ab}({\bf k},x)$
to which the soft fields $A^\mu_a(x)$ couple
via {\it kinetic equations}.
By solving these equations, one can express $\delta N_{ab}$
as a functional of the fields $A^\mu_a$.
The corresponding {\it colour current} 
(with $v^\mu=(1, {\bf k}/k)$) :
\beq\label{jb1}
j_a^\mu(x) \equiv 2g\int\frac{{\rm d}^3k}{(2\pi)^3}\,v^\mu
\,{\rm Tr}\,\Bigl(T^a\delta N({\bf k},x)\Bigr),\eeq
acts as a generating functional for the
(equilibrium) amplitudes of the soft fields \cite{qcd,prept}:
\beq\label{exp0}
j^{a}_\mu \,=\,\Pi_{\mu\nu}^{ab}A_b^\nu
+\frac{1}{2}\, \Gamma_{\mu\nu\rho}^{abc} A_b^\nu A_c^\rho+\,...
\eeq
Here, $\Pi_{\mu\nu}^{ab}=\delta^{ab}\Pi_{\mu\nu}$
is the soft polarization tensor, and the other terms represent
vertex corrections. These are the amplitudes which define the
{\it effective theory} for the soft fields.
When applied to colour excitations at the scale $gT$
\cite{qcd,prept,BIO96}, this strategy provides
 the so-called ``hard thermal loops'' (HTL)
\cite{BP90,FT90,MLB96}. It is our purpose in this paper
to generalize this strategy to the {\it ultrasoft} scale $g^2 T$,
and construct the corresponding amplitudes.

Remarkably, to the order of interest the density matrix
can be parametrized as:
\beq\label{dn0}
\delta N_{ab}({\bf k}, x)\,=\,
 - gW_{ab}(x,{\bf v})\,({\rm d}N_0/{\rm d}k),\eeq
where $N_0(k)\equiv 1/({\rm e}^{\beta k}-1)$
is the Bose-Einstein thermal distribution,
and $W(x,{\bf v})\equiv W_a(x,{\bf v}) T^a$
is a colour matrix in the adjoint representation
which depends upon the velocity ${\bf v}={\bf k}/k$
(a unit vector), but not upon the magnitude $k
=|{\bf k}|$ of the  momentum. Then, the kinetic equations
are written as equations for $W_a(x,{\bf v})$.

Let us briefly
recall the situation at the scale $gT$. The relevant kinetic equation
is a non-Abelian generalization of the Vlasov equation \cite{qcd}:
\beq\label{VLAS}
(v\cdot D_x)^{ab}W_b(x,{\bf v})&=&{\bf v}\cdot{\bf E}^a(x).\eeq
It differs from the corresponding Abelian equation, namely
(with $W(x,{\bf v})$ a fluctuation in the electric charge density)
\beq\label{VLAS0}
(v\cdot \del_x)W(x,{\bf v})&=&{\bf v}\cdot{\bf E}(x),\eeq
merely by the repacement of the ordinary (soft) derivative
$\del_x \sim gT$ by the covariant one $D_x=\del_x+igA$.
Accordingly, the soft gluon polarization tensor derived from 
eq.~(\ref{VLAS}) is formally identical to the photon
polarization tensor obtained from eq.~(\ref{VLAS0}).
In addition, eq.~(\ref{VLAS}) also generates,
through the covariant derivative, an infinite series of
gluon vertices. These are the HTL's alluded to before,
corresponding to one-loop diagrams
with soft external lines and hard internal momenta \cite{BP90,FT90}.
Note that the kinetic equation (\ref{VLAS0}) isolates directly the dominant
contributions of such diagrams, in a gauge invariant manner.

\begin{figure}
\protect \epsfxsize=7.cm{\centerline{\epsfbox{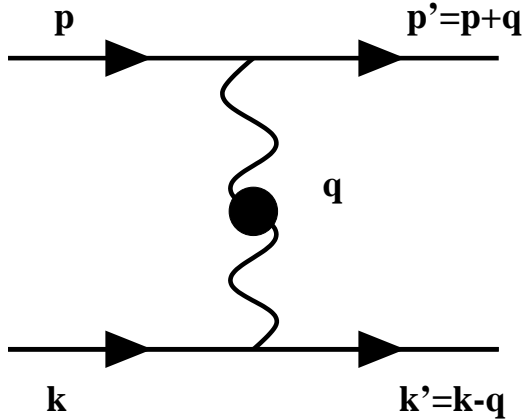}}}
         \caption{Elastic scattering in the (resummed) Born approximation.
The continuous lines refer to hard gluons (these are off-equilibrium
propagators), while the wavy line is the soft gluon exchanged in the
collision. The blob stands for HTL resummation.}
\label{Born}
\end{figure}

This close similitude between the response of Abelian and non-Abelian
plasmas to longwavelength perturbations  
disappears, however, when going to very soft perturbations,
where collisions start to play a role. The effects of the collisions
depend upon the specific excitations one is looking at.
To give a crude estimate of these effects,
one may use the relaxation time approximation, where the kinetic equation is
written as
\beq\label{RTA}
(v\cdot D_x)^{ab}W_b(x,{\bf v})&=&{\bf v}\cdot{\bf E}^a(x)
\,-\,\frac{W^a(x,{\bf v})}{\tau_{col}},\eeq
and $\tau_{col}$ is the typical relaxation time for small 
off-equilibrium colour fluctuations.
Like the damping rate $\gamma$ for hard quasiparticles
\cite{Smilga90,Pisarski93,lifetime,BDV98}, to which it is 
intimately related (see below), the relaxation of colour is dominated
by the singular forward scattering (i.e., by soft momentum transfers
in the collision in Fig.~\ref{Born}), which yields
$\tau_{col} \sim 1/\gamma  \sim 1/(g^2T\ln(1/g))$ 
\cite{Gyulassy93,Heisel94}. Then, eq.~(\ref{RTA}) shows that
the effect of the collisions become a leading order effect
for inhomogeneities at the scale $\del_x \sim g^2 T$, or less.
This should be contrasted with the case of colourless 
fluctuations\footnote{Note also that, for colourless fluctuations,
the analogue of eq.~(\ref{RTA}) will generally involve a
{\it momentum-dependent} relaxation time \cite{Baym97}; 
see Sec. 4.2 below.}, for instance fluctuations
in the momentum or the electric charge distributions,
where the typical relaxation time is much larger,
$\tau_{el} \sim 1/(g^4T\ln(1/g))$,
as it requires large angle scattering \cite{Baym90,Heisel94a,Baym97}.

For colour fluctuations at the scale $g^2 T$,
it is further convenient to constrain the amplitudes
of the associated mean fields such as $|A^\mu_a| \sim gT$;
then the two terms of the {\it ultrasoft}
covariant derivative are of the same order in $g$ (namely
$\del_x \sim gA \sim g^2T$) and, in the derivation of the
kinetic equations, one can consistently preserve
gauge symmetry with respect to the background field \cite{BE}.
There is another reason which makes this
constraint interesting: $|A^\mu_a| \sim gT$ is the typical
amplitude of the thermal fluctuations at the scale $g^2 T$
\cite{SEWM}. These fluctuations have relatively
large amplitudes because of  Bose-Einstein enhancement,
and their dynamics is fully non-linear; as a result,
perturbation theory breaks down at the scale $g^2 T$ \cite{MLB96,ASY96}. Moreover, these 
large amplitude fluctuations make it impossible to give a gauge independent meaning to 
inhomogeneities on scales much larger than $1/g^2T$.
A convenient strategy to deal with this situation is to observe
that the soft modes can be treated as {\it classical fields},
precisely because of their large occupation numbers
\cite{McLerran,ASY96,baryo,Bodeker98} (and references therein).
Then, the non-perturbative dynamics can be studied via
classical lattice simulations of the effective theory for
soft fields \cite{Hu,Moore98,Rajantie99}.

In order to explicitly construct this theory, however, one needs 
to go beyond the relaxation time approximation (\ref{RTA}). 
In fact,  eq.~(\ref{RTA}) is inconsistent with gauge symmetry,
as it leads to a colour current which is not conserved.
The correct kinetic equation, as derived in 
\cite{Bodeker98,BE} (see also Refs. \cite{ASY98,Manuel99,Basag99}),
involves a more complicated collision term, which is local in $x$,
but non-local in ${\bf v}$ (see eq.~(\ref{W1}) below).
 Since this collision term is saturated by soft momentum transfers 
(it is logarithmically sensitive to all momenta $q \simle gT$),
it is useful to isolate the ultrasoft
($q\sim g^2T$) background fields from the soft 
($g^2T \simle q \simle gT$) gluons exchanged in the collisions by
introducing an intermediate scale $\mu$ such as $g^2T\ll\mu\ll gT$ 
(e.g., $\mu\simeq g^2T\ln(1/g)$). Then, the Boltzmann equation 
generates an effective theory for the ultrasoft ($q < \mu$) fields,
corresponding to ``integrating out'' the hard and soft
($q > \mu$) fields to leading  order in perturbation theory
\cite{Bodeker98,BE} (see also Secs. 2 and 3.2 below for a discussion
of the relevant approximations). The scale $\mu$ acts as an 
infrared (IR) cutoff for the collision integral, and as
an ultraviolet (UV) cutoff for the effective theory, and it must
cancel in any complete calculation of ultrasoft correlation
functions (a cancellation referred to as {\it matching}). 

It is our purpose in this paper to study the contribution of 
hard and soft fields to the amplitudes with ultrasoft
external fields ({\it ultrasoft amplitudes} in brief)
by an analysis of the solution to the Boltzmann equation.

In previous applications of the latter
--- namely, to the calculation of the (transverse) 
colour conductivity to leading logarithmic accuracy
\cite{Gyulassy93,Heisel94,Bodeker98,ASY98} ---, the
non-local piece of the collision term turned out not to be important.
But this was specific to that particular approximation,
which has ignored all the non-local and non-linear effects in
the problem (essentially because the drift term $v\cdot D_x \sim g^2T$ 
has been neglected as compared to $\gamma \sim g^2T\ln(1/g)$;
see Sec. 3.4  for more details). In that situation,
the ultrasoft amplitudes in eq.~(\ref{exp0}) collapsed
to a single, local quantity, namely the colour conductivity.

Our intention here is to go
beyond this leading logarithmic approximation and study the
generic ultrasoft amplitudes generated by the Boltzmann equation
for colour relaxation. This includes, in particular, the
non-local effects in space and time, as governed by the drift term
$v\cdot D_x$, and also the non-local effects in ${\bf v}$
coming from the collision term.
Because of the latter, the Boltzmann equation
cannot be exactly solved in general. Thus, we will not be able
to provide full expressions for the ultrasoft amplitudes
except in some simple limits (cf. Sec. 4.2).
Still, many of the important properties
of these amplitudes can be inferred from an analysis of the
Boltzmann equation (cf. Secs. 3.1, 3.3 and 4 below).
Moreover, some formal solutions can also
be obtained, by iterations, and this is specially
useful for comparison with diagrammatic perturbation theory
(cf. Secs. 3.2 and 3.3). The Boltzmann equation
accomplishes a non-trivial resummation of the perturbative
expansion, made evident by the derivation of this
equation from quantum field theory \cite{BE}, where the collision 
term has a direct diagrammatic interpretation (see Sec. 2 below
for a short review of this derivation). 

Let us briefly enumerate here the main properties of the 
ultrasoft amplitudes, to be derived below in this paper:
For generic external momenta of order $g^2 T$, these amplitudes
are of the same order in $g$ as the HTL's, which they
generalize by taking into account the effects of the collisions.
The ultrasoft amplitudes share many of the remarkable
properties of the HTL's: {\it i}) they are non-perturbative,
in the sense that in the kinematical regime of interest
($\omega\ll p\simle g^2T$), they are as large as the corresponding
tree-level amplitudes; {\it ii}) they are gauge-fixing 
independent, as they involve only collisions among on-shell,
hard, transverse gluons, {\it iii}) they satisfy simple Ward
identities, which express the conservation of the colour current;
{\it iv}) in the static limit $\omega \to 0$, they reduce to the 
usual Debye mass term $m_D^2= g^2 NT^2/3$ for the electric fields.
That is, to the order of interest, collisions do not modify
Debye screening (see also Ref. \cite{ASY99}, and Sec. 3.3 below
for more details).

On the other hand, there are also significant differences
with respect to the HTL's: {\it i}) The ultrasoft amplitudes
have no Abelian counterpart: in QED, the effects of the
collisions become important only at the scale $e^4 T$ (see also
Sec. 2 below).
{\it ii}) Unlike the HTL's, which correspond to one-loop Feynman
graphs, the ultrasoft amplitudes receive contributions from an infinite
series of multi-loop diagrams, with a specific structure
(essentially, chains of ladder diagrams). Recently, the first few
such diagrams for the polarization tensor have been explicitly
computed by B\"odeker \cite{Bodeker99}, with results which
agree with the (first order iteration of the) solution to the
Boltzmann equation. But it is clear that, in general,
computing directly these diagrams would be a tedious exercice,
especially since important cancellations occur among the various
graphs \cite{BE,Bodeker99}; this will be further discussed in Secs.
3.2 and 4.1 below. {\it iii}) The resummation of the collision effects
drastically modify the longwavelength behaviour of the
transverse colour conductivity $\sigma_T(\omega=0,p\to 0)$:
the would-be divergence of the HTL  result for $\sigma_T$,
namely $\sigma_T^{(0)} \propto m_D^2/p$, is now screened away
by $\gamma$, with the net result that $\sigma_T(\omega=0,p\to 0)= 
m_D^2/3(\gamma - \delta)$. Here $\delta$ is a  term of order 
$g^2 T$ (and which satisfies $|\delta|<\gamma$), to be computed
in Sec. 4.2 below. {\it iv}) For ultrasoft momenta $p\sim g^2 T$,
the above formula for  $\sigma_T$ holds  up to corrections
of ${\rm O}(\ln^{-2})$ (with $\ln\equiv \ln(1/g)$).

\setcounter{equation}{0}
\section{The Boltzmann equation}

In this section, we review the main features of the Boltzmann
equation for colour relaxation which was recently derived in Ref. \cite{BE}.
There are no new results to be reported here, but the equations
derived in \cite{BE} will be presented in a slightly different way,
to better emphasize the difference between 
coloured and colourless excitations,
and also between Abelian and non-Abelian gauge theories.
Moreover, the diagrammatic representation of the collision term
will be explained in more detail, in order to facilitate the
discussion of the diagrammatic interpretation of the ultrasoft
amplitudes, in Sec. 3.2 below.

The space-time inhomogeneities in the distribution of
the hard particles (transverse gluons) are described by a 
density matrix $\delta\acute G_{ab}(k,x)$ where the momentum $k$
is hard ($k\sim T$) and on-shell ($k_0^2={\bf k}^2$), and the
derivative $\del_x$ is ultrasoft ($\del_x\sim g^2T$).
Below, we shall be interested either in colourless fluctuations
(in which case $\delta\acute G_{ab} =\delta_{ab}\delta \acute G$),
or in coloured ones in the adjoint representation
(such as $\delta\acute G_{ab} = if^{abc}\delta \acute G_c$). Also, we shall
find convenient to use the following parametrization for the density 
matrix, where the on-shell structure is explicit:
\beq\label{GNW}
\delta\acute G_{ab}(k,x)&\equiv&
-\rho_0(k) W_{ab}(k,x)\,\frac{{\rm d}N_0}{{\rm d}k_0}\nonumber\\
&=&\beta \rho_0(k) W_{ab}(k,x)\,N_0(k_0)[1+N_0(k_0)].\eeq 
In this equation, $N_0(k_0) = 1/({\rm e}^{\beta k_0}-1)$ and
$\rho_0(k)=2\pi \epsilon(k_0)\delta(k^2)$ are, respectively, the
thermal distribution and the spectral density for hard
transverse gluons, and the new function $W_{ab}(k,x)$ has
support only at the mass-shell:
$W_{ab}(k,x)= \theta(k_0) W_{ab}({\bf k},x)+\theta(-k_0)W_{ba}
(-{\bf k},x)$. [The density matrix
$\delta N_{ab}({\bf k}, x)$ in the Introduction, eq.~(\ref{dn0}),
is related to the function $W_{ab}({\bf k},x)$ by $\delta N_{ab}({\bf k}, x)
= -W_{ab}({\bf k},x)({\rm d}N_0/{\rm d}k)$.]

The Boltzmann equation is the kinetic equation satisfied by
the density matrix to leading order in $g$ \cite{BE}. It reads
(in matrix notations):
 \beq\label{bolt}
2\Bigl[k\cdot D_x,\,\delta\acute G(k,x)\Bigr]
\,-\,2gk^\mu  F_{\mu\nu}(x)\del^\nu_k G_0^<(k)\,=\,
C(k,x).\eeq
In the l.h.s.,  $k\cdot D_x$ is the
gauge-covariant drift operator, with $D^\mu\equiv\del^\mu+igA^\mu$
and $\del_x\sim gA\sim g^2T$, so that $D_x={\rm O}(g^2T)$;
$k^\mu  F_{\mu\nu}(x)\del^\nu_k$,
with $F_{\mu\nu}\equiv [D_\mu,D_\nu]/ig$, is the ``force'' term 
acting on the equilibrium correlation function $G^<_0(k)$:
 \beq\label{G0}
G^<_0(k)\,\equiv\, \rho_0(k)N_0(k_0),
\qquad G^>_0(k)\equiv \rho_0(k)[1+N_0(k_0)].\eeq
(The second function $G^>_0(k)$ will be needed below.)

In the r.h.s. of eq.~(\ref{bolt}), $C(k,x)$
is the collision term associated to the one-gluon exchange process
depicted in Fig.~\ref{Born}. A priori, all the lines in this
figure (that is, both the external lines associated
with the colliding particles, and the wavy line associated
to the exchanged gluon) are off-equilibrium propagators. However,
to the order of interest, the collision term can be linearized
with respect to the off-equilibrium fluctuations in the propagators
of the external lines, and the internal propagator can be taken
to be the equilibrium propagator. Since the collision term for 
colour relaxation is dominated by
soft momentum transfers ($g^2T \simle q \simle gT$) 
\cite{Gyulassy93,Heisel94,Bodeker98,ASY98,BE}, the
propagator of the exchanged gluon
has to be dressed with the
corresponding hard thermal loop \cite{BIO96,MLB96}.

\begin{figure}
\protect \epsfxsize=11.cm{\centerline{\epsfbox{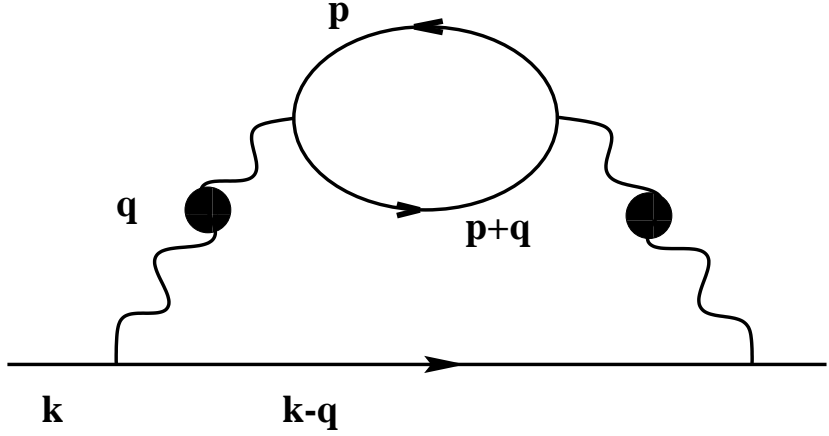}}}
         \caption{Self-energy describing collisions
in the (resummed) Born approximation. All the lines represent
off-equilibrium propagators. The continuous lines refer to the
hard colliding particles in Fig.~\ref{Born}. The wavy lines
with a blob denote soft
gluon propagators dressed by the screening effects.}
\label{S11}
\end{figure}

The scattering 
process in Fig.~\ref{Born} can be associated to the collisional
self-energy in Fig.~\ref{S11} (see Ref. \cite{BE}
for more details on the formalism). Upon linearization, 
this leads to the four processes displayed in
Fig.~\ref{FLS}. Each diagram involves fluctuations in one of
the four external lines in  Fig.~\ref{Born}. Thus,
$C=C_1+C_2+C_3+C_4$, where $C_1(k,x)$ involves
the fluctuations $\delta\acute G(k,x)$ in the incoming field with
momentum ${\bf k}$ (Fig.~\ref{FLS}.a), and
$C_2$, $C_3$ and $C_4$ involve fluctuations
along the lines with momenta ${\bf k}'$, ${\bf p}$ and ${\bf p}'$
(Figs.~\ref{FLS}.b, c and d, respectively).
In these figures, the off-equilibrium propagators are marked with
a cross; all the other lines denote equilibrium propagators.
In particular, $C_1(k,x)= -\Gamma(k)\,\delta\acute G(k,x)$, where
\beq\label{EQG}
\Gamma(k)\equiv \Sigma^<_{eq}(k)
 - \Sigma^>_{eq}(k)= -2\,{\rm Im} \Sigma_R(k)
\,,\eeq
is the quantity which determines the quasiparticle
damping rate $\gamma\equiv \Gamma(k_0=k)/(4k)\,\sim\,
g^2T\ln(1/g)$ \cite{BP90,Smilga90,Pisarski93,lifetime}.

By using the parametrization (\ref{GNW}) for the density matrix,
the (linearized) collision term can be compactly written 
as\footnote{Eq.~(\ref{LINCOL}) is merely a convenient
rewriting of  eqs.~(3.99)--(3.104) in Ref. \cite{BE}.}:
\begin{figure}
\protect \epsfxsize=14.5cm{\centerline{\epsfbox{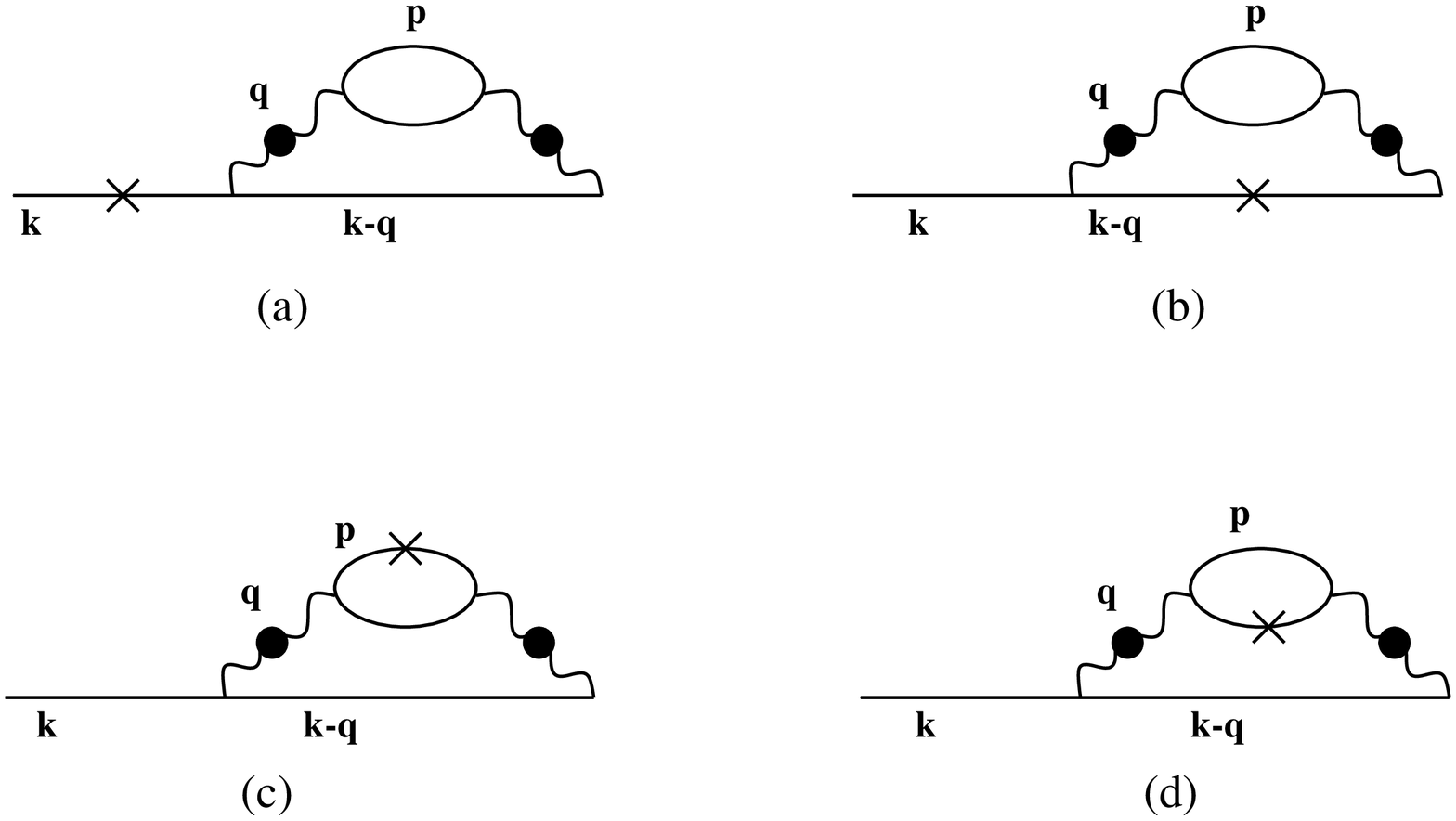}}}
         \caption{Pictorial representation of the linearized collision term.
Each one of the four diagrams correspond to off-equilibrium fluctuations in one
of the colliding fields (the one which is marked with a cross).
All the unmarked propagators are in equilibrium.}
\label{FLS}
\end{figure} 
\beq\label{LINCOL}
C_{ab}(k,x)&=&
-\int {\rm d}{\cal T}\left| {\cal M}_{pk\to p'k'}\right|^2
\,N_0(k_0)\,N_0(p_0)\,
[1+N_0(k^\prime_0)]\,[1+N_0(p^\prime_0)]\nonumber\\&{}&\qquad\times
\Bigl\{N\Bigl(NW_{ab}(k,x) - (T^aT^b)_{cd}
W_{cd}(k^\prime,x)\Bigr)\,+\nonumber\\&{}&\qquad\,\,\,\,\,\,\,\,+\,
(T^aT^b)_{c\bar c}(T^cT^{\bar c})_{d\bar d}\Bigl(W_{\bar d d}(p,x)
-W_{d\bar d}(p',x)\Bigr)\Bigr\}.\eeq
In this equation, $\left| {\cal M}_{pk\to p'k'}\right|^2$
is the matrix element squared corresponding to the 
one-gluon exchange depicted in Fig.~\ref{Born},
and  ${\rm d}{\cal T}$ is a compact notation for the measure of the
phase-space integral:
\beq\label{PHI}
\int {\rm d}{\cal T}\,\equiv\,\beta
\int\frac{{\rm d}^4p}{(2\pi)^4} \int\frac{ {\rm d}^4 q}{(2\pi)^4}\,
 \rho_0(k)\rho_0(p)\rho_0(p+q)\rho_0(k-q).\eeq
The four terms within the braces in eq.~(\ref{LINCOL})
are in one to one correspondance with the diagrams
\ref{FLS}.a, b, c and d.
The appearance of the matrix element squared, and also of 
the various equilibrium statistical factors in eq.~(\ref{LINCOL}),
is familiar. What is specific to the problem at hand 
is the colour structure in eq.~(\ref{LINCOL}), 
which is at the origin of an important
difference between coloured and colourless excitations:

Consider first the case of a colourless fluctuation,
for which $\delta\acute G_{ab} =\delta_{ab}\delta \acute G$, and
$W_{ab}=\delta_{ab} W$. The various colour traces
in eq.~(\ref{LINCOL}) are trivial (e.g., $(T^aT^b)_{cc}=
N\delta_{ab}$), so that $C_{ab}=\delta_{ab}C$, with
\beq\label{CLESS}
C(k,x)&=&
-N^2\int {\rm d}{\cal T}\left| {\cal M}_{pk\to p'k'}\right|^2
\,N_0(k_0)\,N_0(p_0)\,[1+N_0(k^\prime_0)]\,[1+N_0(p^\prime_0)]
\nonumber\\&{}&\qquad\times
\Bigl\{W(k,x)-W(k^\prime,x)+W(p,x)-W(p^\prime,x)\Bigr\}.\eeq
This is the standard collision term for one-gluon exchange
used in previous applications of kinetic theory to the
hot quark-gluon plasma 
or to the electroweak plasma \cite{Baym97}.

What is remarkable about eq.~(\ref{CLESS}) is that the
corresponding phase-space integral is dominated by relatively
hard momentum transfers $gT\simle q\simle T$, even though each of the
four individual terms in the r.h.s. is actually saturated by
soft momenta. This is a consequence of
the cancellation of the leading infrared contributions among the
various terms \cite{BE}. For instance,  for soft $q$, $W(k',x)\equiv W(k-q,x)
\approx W(k,x)$, so that the IR contributions to the first two terms
in eq.~(\ref{CLESS}) cancel each other. This corresponds to a cancellation
among the graphs displayed in Figs.~\ref{FLS}.a and b, to
be further discussed in Sec. 3.2 below.
A similar cancellation occurs between the last two
terms in eq.~(\ref{CLESS}), namely $W(p,x)$ and $W(p',x)$.
Thus, in order to see the leading IR ($q\ll T$) behaviour
of the full integrand in eq.~(\ref{LINCOL}), one has to expand
$W(k',x)$ and $W(p',x)$ to higher orders in $q$. This 
generates extra factors of $q$ which remove the most severe
IR divergences in the collision integral.
(This is the familiar $(1-\cos\theta)$ factor of
the ``transport cross section'' \cite{PhysKin}.) As a result, the typical
rate involved in the calculation of the transport coefficients
like the shear viscosity is $\sim\, g^4T\ln(1/g)$,
where the logarithm
originates from screening effects at the scale $gT$
\cite{Baym90,Heisel94a}. This is suppressed
by one power of $\alpha \equiv g^2 /4\pi$ with respect to the 
damping rate $\gamma\,\sim\, g^2T\ln(1/g)$.

Consider now the case of colour fluctuations
corresponding to a density matrix
$W(k,x)$ of the form $W(k,x)\equiv W_a(k,x)T^a$.
The colour algebra
in eq.~(\ref{LINCOL}) can be performed with the following identities:
\beq\label{TRACE}
{\rm Tr} (T^aT^bT^c) \,=\,if^{abc}\,\frac{N}{2}\,,\qquad
(T^aT^b)_{c\bar c}(T^cT^{\bar c})_{d\bar d}(T^e)_{\bar d d}
\,=\,if^{abe}\,\frac{N^2}{4}\,.\eeq
The resulting collision term is of the form $C=C_a T^a$ with
\beq\label{CCOL}
C_a(k,x)&=& -N^2\int {\rm d}{\cal T}\left| {\cal M}_{pk\to p'k'}\right|^2
\,N_0(k_0)\,N_0(p_0)\,[1+N_0(k^\prime_0)]\,[1+N_0(p^\prime_0)]
\nonumber\\&{}&\qquad\times
\left\{W_a(k,x)-\,\frac{1}{2}\,
W_a(k^\prime,x)-\,\frac{1}{4}\Bigl(W_a(p,x)+W_a(p^\prime,x)
\Bigr)\right\}.\eeq
There are two notable differences with respect to eq.~(\ref{CLESS}):\\
{\it i}) The first two terms within the braces enter with a relative
factor 1/2, so they do not cancel each other when $q\to 0$ :
\beq\label{SIMP1} 
W_a(k,x)-\,\frac{1}{2}\,W_a(k-q,x)\,\simeq\,\,\frac{1}{2}\,W_a(k,x).\eeq
Rather, their overall contribution is
{\it half} the corresponding contribution of the first term alone,
that is, $\Gamma(k)/2$.\\
{\it ii}) The last two terms in eq.~(\ref{CCOL}) enter with a
factor $1/4$ and {\it add} each other. This is so because
the colour matrix $W_{\bar d d}(p,x)$  in eq.~(\ref{LINCOL})
is antisymmetric, rather than symmetric, as it is for colourless
fluctuations. Accordingly, for soft $q$,
\beq\label{SIMP2} \frac{1}{4}\Bigl(W_a(p,x)+W_a(p+q,x)\Bigr)
\,\simeq\,\,\frac{1}{2}\,W_a(p,x).\eeq
Thus, for colour fluctuations, the colour structure of the
collision term prevents a complete cancellation of the leading
infrared contributions: like the damping rate, the collision term 
for colour relaxation is saturated by soft momentum transfers
($g^2T \simle q\simle gT)$, for which eqs.~(\ref{SIMP1})
and (\ref{SIMP2}) hold and the collision term (\ref{CCOL})
simplifies to \cite{ASY98,BE}:
\beq\label{CSIMP}
C_a(k,x)\simeq -\,\frac{N^2}{2}
\int {\rm d}{\cal T}\left| {\cal M}_{pk\to p'k'}\right|^2
\,\frac{{\rm d}N_0}{{\rm d}k_0}\,\frac{{\rm d}N_0}{{\rm d}p_0}\,
\left\{W_a(k,x)\,-\,W_a(p,x)\right\}.\eeq
In the same approximation, the matrix element $|{\cal M}|^2$ 
can be evaluated as:
\beq\label{COLM1}
|{\cal M}|^2\,=\,16g^4\varepsilon_k^2\varepsilon_p^2\,
 \Big|{}^*{\cal D}_l(q)
+ ({\bf \hat q \times v})\cdot ({\bf \hat q \times v}^\prime)\,
{}^*{\cal D}_t(q)\Big|^2,\eeq
where ${\bf v}\equiv \hat{\bf k}$, ${\bf v}^\prime \equiv \hat{\bf p}$, 
and ${}^*{\cal D}_l$ and ${}^*{\cal D}_t$ are the longitudinal 
(or electric) and the transverse (or magnetic) components of the
(retarded) gluon propagator, 
in the hard thermal loop approximation \cite{BIO96,MLB96}.
The phase-space measure (\ref{PHI}) can be similarly simplified.
This eventually yields a simpler equation
for the density matrix $W_a({\bf k},x)$
which, remarkably, is consistent with $W_a({\bf k},x)$ being
independent of the magnitude $k\equiv |{\bf k}|$ of the hard momentum.
That is,
\beq
W_a({\bf k},x)&\equiv & gW_a(x,{\bf v}),\eeq where
${\bf v}\equiv {\bf k}/k$ is the velocity of the termal particle
(a unit vector), and a factor
of $g$ has been introduced to keep in line
with the notations of Ref. \cite{BE}.

Finally.
the Boltzmann equation, written as an equation for
$W_a(x,{\bf v})$, reads \cite{BE}:
\beq\label{W1}
(v\cdot D_x)^{ab}W_b(x,{\bf v})&=&{\bf v}\cdot{\bf E}^a(x)-m_D^2
\frac{g^2 N T}{2}\int\frac{{\rm d}\Omega'}{4\pi}
\,\Phi({\bf v\cdot v}^\prime)\Bigl\{W^a(x,{\bf v})-
W^a(x,{\bf v}^\prime)\Bigr\}.\nonumber\\&{}&\eeq
The angular integral above runs over all the directions of the unit
vector ${\bf v}^\prime$, and $m_D^2$ is the Debye mass squared:
\beq\label{MD}
m^2_D\,\equiv\,-\,\frac{g^2 N}{\pi^2}\int_{0}^\infty 
{\rm d}p \,p^2\,\frac{{\rm d}N_0}{{\rm d}p}\,=\,\frac{g^2N T^2}{3}\,.\eeq
Furthermore:
\beq\label{PHII}\Phi({\bf v\cdot v}^\prime)\equiv(2\pi)^2
\int\frac{{\rm d}^4 q}{(2\pi)^4}\,
\delta(q_0- {\bf q\cdot v})
\delta(q_0- {\bf q\cdot v}^\prime)
\Big|{}^*{\cal D}_l(q)+ ({\bf v}_t\cdot{\bf v}_t^\prime)\,
{}^*{\cal D}_t(q)\Big|^2,\,\,\,\eeq
with the two delta functions expressing the energy conservation
at the two vertices of the scattering process in Fig.~\ref{Born}.
 Up to a normalization, the function $\Phi({\bf v\cdot v}^\prime)$
represents the cross section for the  collision  between
two hard particles with velocities ${\bf v}$ and ${\bf v}'$ exchanging
(in  the $t$-channel) a soft (dressed) gluon. 

The collision term in  eq.~(\ref{W1}) involves two pieces:
one which is local in ${\bf v}$ (proportional to
$W^a(x,{\bf v})$), and one which is non-local
(involving the kernel
$\Phi({\bf v\cdot v}^\prime)$. The coefficient
of the local piece is proportional to $\Gamma$ :
\beq\label{gamma1}
m_D^2\,\frac{g^2 N T}{2}\int\frac{{\rm d}\Omega'}{4\pi}
\,\Phi({\bf v\cdot v}^\prime)\,=\,\frac{\Gamma(k_0=k)}{4k}\,
\equiv\,\gamma.\eeq
By using the expression above, we can rewrite the
Boltzmann equation (\ref{W1}) in the following way:
\beq\label{W10}
(v\cdot D_x)^{ab}W_b(x,{\bf v})&=&{\bf v}\cdot{\bf E}^a(x)-
\gamma\Bigl\{W^a(x,{\bf v})\,-\,\langle W^a(x,{\bf v})\rangle
\Bigr\},\eeq
which emphasizes the fact that the quasiparticle damping rate
$\gamma$ sets the time scale for colour relaxation:
 $\tau_{col}\sim 1/\gamma \sim 1/(g^2T\ln(1/g))$ 
\cite{Gyulassy93} (see also Sec. 4.2 below).
In eq.~(\ref{W10}) we have introduced a notation which will
be used hereafter: for an arbitrary function
of ${\bf v}$, say $F({\bf v})$, we denote by
$\langle F({\bf v})\rangle$ its angular average with weight
function $\Phi({\bf v\cdot v}^\prime)$:
\beq\label{AVF}
\langle F({\bf v})\rangle\,\equiv\,\frac{
\int\frac{{\rm d}\Omega'}{4\pi}\,\Phi({\bf v\cdot v}^\prime)
F({\bf v}')}{\int\frac{{\rm d}\Omega'}{4\pi}\,
\Phi({\bf v\cdot v}^\prime)}\,,\eeq
which is still a function of ${\bf v}$. 

We conclude this section by recalling that eq.~(\ref{W1})
is invariant under the gauge transformations of the
{\it background field}, and also with respect to
the choice of a gauge for the shortwavelength fluctuations 
(here, the hard ($k\sim T$) fields which take part in the collective
motion and the soft ($g^2 T\simle q\simle gT$) gluons
which are exchanged in the collision process). In Ref. \cite{BE},
eq.~(\ref{W1}) was derived in Coulomb gauge, but we expect it
to be gauge-fixing independent. Except for the collision term, this has
been explicitly verified in \cite{qcd} (see also Refs. \cite{BP90,FT90}).
The collision term should be gauge-fixing independent as well,
since it involves only the off-equilibrium fluctuations of the
(hard) transverse gluons, together with the (gauge-independent)
matrix element squared (\ref{COLM1}). However, an explicit proof
comparable to the corresponding one for the
non-Abelian Vlasov equation \cite{qcd} is somewhat tedious:
in an arbitrary gauge (e.g., a covariant one), one
has to consider collisions involving fictitious degrees of freedom
(hard longitudinal gluons and ghosts), and verify that their respective
contributions to the collision term mutually cancell.

\setcounter{equation}{0}
\section{Ultrasoft amplitudes}
In this section, we introduce and study the ultrasoft amplitudes,
i.e., the contributions to the one-particle irreducible amplitudes
with ultrasoft external lines which are obtained from
the solution to the Boltzmann equation.

\subsection{The induced current}

The longwavelength colour fluctuations of the hard particles
generate a colour current given by (the factor of 2 below stands for
the two transverse polarizations):
\beq\label{jb}
j^a_\mu(x)=2g\int\frac{{\rm d}^4k}{(2\pi)^4}\,k_\mu
\,{\rm Tr}\,\Bigl(T^a\delta\acute G(k,x)\Bigr),\eeq
which acts as a source in the Yang-Mills equations for the
ultrasoft colour fields $A^\mu_a$:
\beq\label{avA}
 (D^\nu F_{\nu\mu})^a(x)&=&j_\mu^a(x).
\eeq 
By using the parametrization (\ref{GNW}) for the density matrix,
one can perform the integral over the radial momentum $k\equiv
|{\bf k}|$ to obtain:
\beq\label{j1}
j^\mu_a(x)&=&m_D^2\int\frac{{\rm d}\Omega}{4\pi}
\,v^\mu\,W_a(x,{\bf v}),\eeq
with the Debye mass $m_D$ defined in eq.~(\ref{MD}).
By using the equation of motion (\ref{W1}) for $W^a(x,{\bf v})$,
one can verify that the current (\ref{j1}) is covariantly
conserved, 
\beq\label{COVCONS}
D_\mu j^\mu\,=\,0,\eeq
as necessary for
the consistency of the mean field equations of motion (\ref{avA})
(recall that $D^\mu D^\nu F_{\nu\mu}=0$).
Indeed, eq.~(\ref{W1}) implies:
\beq\label{VERIF} D_\mu j^\mu&=&m_D^2\int\frac{{\rm d}\Omega}{4\pi}\,
{\bf v}\cdot{\bf E}^a(x)\nonumber\\
&-&m_D^4\frac{g^2 N T}{2}\int\frac{{\rm d}\Omega}{4\pi}
\int\frac{{\rm d}\Omega'}{4\pi}
\,\Phi({\bf v\cdot v}^\prime)\Bigl\{W^a(x,{\bf v})-
W^a(x,{\bf v}^\prime)\Bigr\},\eeq
which is zero because both terms in the r.h.s. vanish
after the angular integration. 

By solving the Boltzmann equation,
one can obtain the density matrix $W_a(x,{\bf v})$, and therefore also
the induced current $j^\mu_a(x)$, as functionals of the gauge fields 
$A^\mu_a(x)$. Since eq.~(\ref{W1}) is non-linear with respect to
the fields $A^\mu_a$, the resulting functional $j_a^\mu[A]$ will be 
non-linear as well, and can be formally expanded as follows:
\beq\label{exp}
j^{a}_\mu \,=\,\Pi_{\mu\nu}^{ab}A_b^\nu
+\frac{1}{2}\, \Gamma_{\mu\nu\rho}^{abc} A_b^\nu A_c^\rho+\,...
\eeq
The coefficients in this expansion are the one-particle-irreducible 
amplitudes of the fields $A^\mu_a$, evaluated in thermal equilibrium
\cite{BIO96,prept,qcd}.
For instance, $\Pi_{\mu\nu}^{ab}=\delta^{ab}\Pi_{\mu\nu}$
is the polarization tensor, $\Gamma_{\mu\nu\rho}^{abc}\equiv if^{abc}
\Gamma_{\mu\nu\rho}$ is a correction to the 3-gluon vertex, etc.
These amplitudes will be referred to as
the {\it ultrasoft amplitudes}. 

Some of the properties of the ultrasoft amplitudes follow
immediately from the previous discussion:
For generic momenta of order $g^2 T$,
they are of the same order in $g$ as the hard thermal
loops \cite{BP90,FT90,qcd,BIO96,MLB96,prept}, which they
generalize by including the effects of the collisions.
Furthermore, they are gauge-fixing independent (like the Boltzmann equation
itself), indicating  that only the {\it physical}
hard degrees of freedom of the plasma (namely, the on-shell 
transverse gluons) contribute to these amplitudes.
Also, they satisfy simple Ward identities which follow from the 
conservation law (\ref{COVCONS}) by successive differentiations with 
respect to the fields $A^\mu_a$. For instance:
\beq\label{Wtr}
P^\mu\,\Pi_{\mu\nu}(P)&=&0,\nonumber\\
P^\mu\Gamma_{\mu\nu\rho}(P,Q,R)&=&\Pi_{\nu\rho}
(Q)-\Pi_{\nu\rho}(R)\,.\eeq
All these properties are, of course,
very reminiscent of the hard thermal loops, and, as we shall
see later, there are other similarities. But let us
first discuss the interpretation of the ultrasoft amplitudes
in terms of Feynman diagrams.

\subsection{Diagrammatic interpretation of the ultrasoft amplitudes}

In this subsection, we discuss the interpretation of the solution to
the Boltzmann equation in terms of Feynman diagrams. (See also Refs.
\cite{Jeon93,JY96} for a related analysis in the context of scalar
field theory, and Ref. \cite{Bodeker99} for a recent calculation
of some of the diagrams relevant to QCD, namely those in  Fig.~\ref{PI2} below.)
This analysis will show that, unlike the HTL's --- 
which correspond to one-loop diagrams \cite{BP90,FT90} ---, 
the ultrasoft amplitudes receive contributions from an infinite
set of multi-loop Feynman graphs, which, in the kinematical
regime of interest, contribute all at the same order in $g$.

Our discussion here will be only qualitative: we shall not compute
Feynman graphs explicitly, but rather rely on the diagrammatic
representation of the collision term (cf. Figs.~\ref{Born},~\ref{S11}
and~\ref{FLS}) in order to identify, by iterations, the structure
of the diagrams contributing to the ultrasoft amplitudes. 

To carry out the analysis, it is convenient
to use the original form of the Boltzmann equation, where the collision 
term is directly related to the self-energy. This is eq.~(\ref{bolt}) 
with the collision term (\ref{LINCOL}), which we rewrite here as follows:
 \beq\label{B}
\Bigl[v \cdot D_x,\,\delta\acute G(k,x)\Bigr]\,=\,
gv^\mu  F_{\mu\nu}(x)\del^\nu_k G_0^<(k)\,+\,C(k,x),
\nonumber\\
C=C_1+C_2+C_3+C_4,\qquad C_1\,=\,-\,\frac{1}{2k_0}\,\Gamma(k)
\,\delta\acute G(k,x),\eeq
where $v^\mu=k^\mu/k_0$, and
the four pieces of the collision term correspond,
respectively, to the linearized fluctuations depicted in 
Figs.~\ref{FLS}.a, b, c and d. (Note that the present normalization
of the collision term differs by a factor $1/2k_0$ from the previous one
in eq.~(\ref{bolt}).) For comparison
with perturbation theory, it is useful to regard the collision term
as a ``small perturbation'' and solve the Boltzmann equation
(\ref{B}) formally by iterations.

\begin{figure}
\protect \epsfxsize=10.5cm{\centerline{\epsfbox{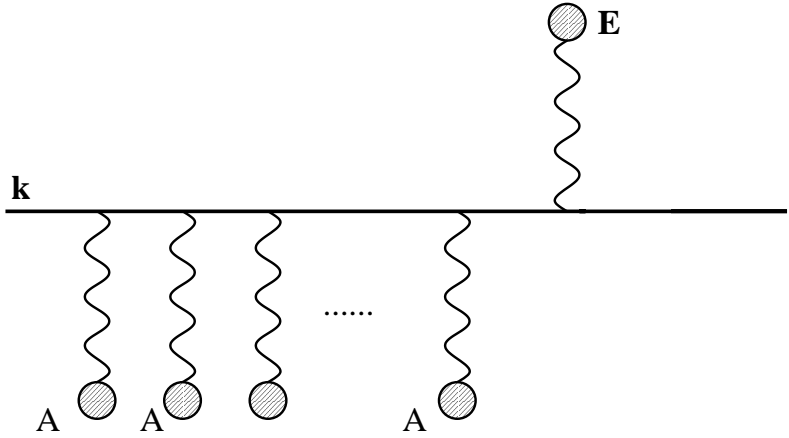}}}
         \caption{Mean field approximation, or zeroth order iteration
for the solution to the Boltzmann equation, eq.~(\ref{FL0}).}
\label{D}
\end{figure}
The zeroth order iteration is the solution to eq.~(\ref{B})
with the collision terms excluded:
\beq\label{FL0}
\delta \acute G^{(0)}\,=\,g\,\frac{1}{v\cdot D}
\,v^\mu  F_{\mu\nu}\del^\nu G_0^<\,.\eeq
Here, $1/(v\cdot D)$ is a compact, but formal, notation for the
retarded Green's function $\Delta_R(x,y;{\bf v})$
of the covariant drift operator $v\cdot D$. This satisfies:
\beq\label{Gret}
(v\cdot D_x)_{ab}\Delta_R^{bc}(x,y;{\bf v})\,=\,\delta^{ac}
\delta^{(4)}(x-y),\eeq
with $\Delta_R(x,y;{\bf v})=0$ for $x_0< y_0$, and has
the following expression (with $t \equiv x_0-y_0$):
\beq\label{GR}
\Delta_{R}^{ab}(x,y;{\bf v})\,=\,\theta (t)\,\delta^{(3)}
\Bigl({{\bf x}}-{{\bf y}}-{\bf v}t
\Bigr )\,U^{ab}(x,y)\,\equiv\,\Bigl\langle x,\,a\Bigl|\frac{i}
{i(v\cdot D)+i\epsilon}\Bigr|y,\,b\Bigr\rangle
,\eeq
where $U(x,y)$ is the Wilson line connecting the points
$x$ and $y\,$:
\beq\label{U}
U(x,y)={\rm e}^{-ie\int {\rm d}z^\mu A_\mu(z)},
\eeq
and the integration path in eq.~(\ref{U})
is fixed by the delta function in eq.~(\ref{GR}). 
$\Delta_R(x,y;{\bf v})$ is the eikonal propagator along the
straightline trajectory of velocity ${\bf v}$. 

\begin{figure}
\protect \epsfxsize=13.cm{\centerline{\epsfbox{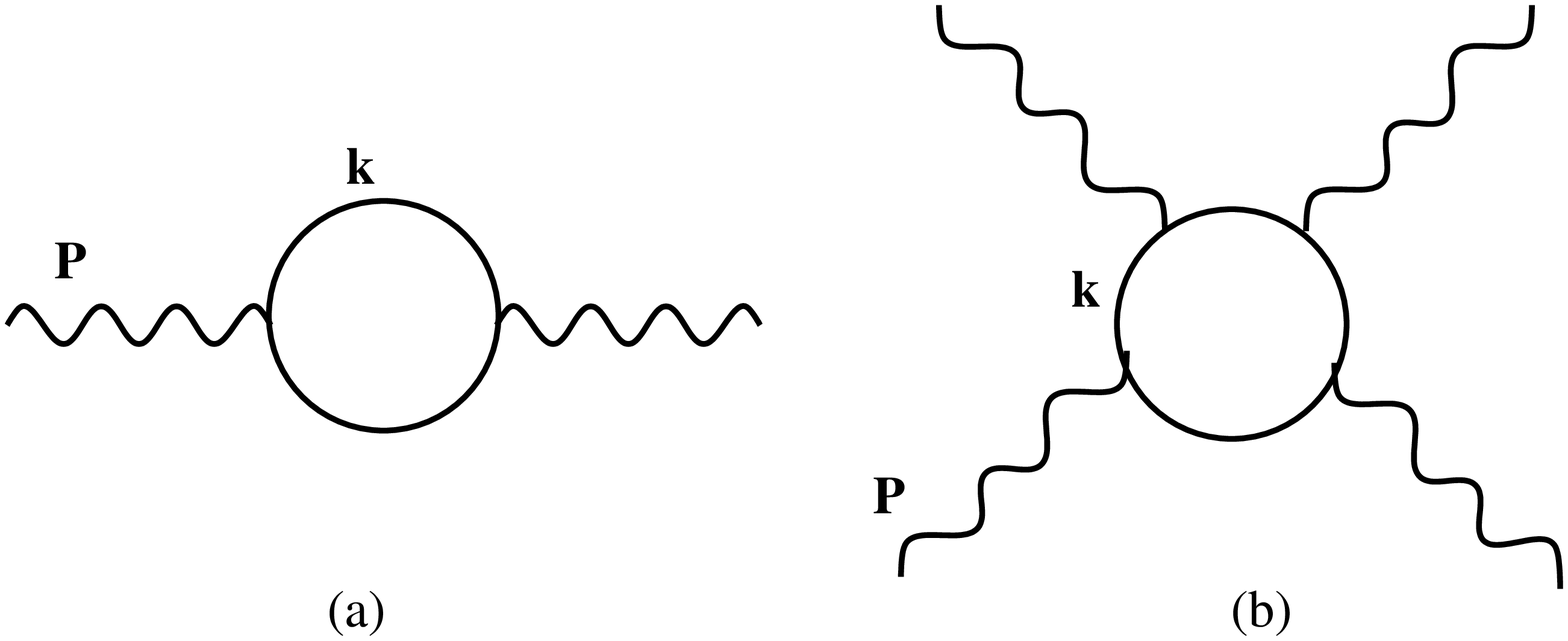}}}
         \caption{Two and four-gluon vertices in the HTL approximation.}
\label{C}
\end{figure}

Eq.~(\ref{FL0}) can be given the diagrammatic representation in 
Fig.~\ref{D} where, for more clarity, we have distinguished the 
insertion of the electric mean field ${\bf E}_a$, the
``Lorentz force'' in the r.h.s. of eq.~(\ref{B}), from
the insertions of colour fields $A^\mu_a$ due to
the covariant derivative $v\cdot D$ in the l.h.s. Thus, the
propagator on the left of the electric field ${\bf E}_a$ is
the eikonal propagator (\ref{GR}), while the propagator on
the right is $\del_k G^<_0(k)$.
The corresponding colour current, namely:
\beq\label{jb0}
j^{(0)\,a}_\mu(x)=2g\int\frac{{\rm d}^4k}{(2\pi)^4}\,k_\mu
\,{\rm Tr}\,\Bigl(T^a\delta\acute G^{(0)}(k,x)\Bigr)\,=\,
m_D^2\int\frac{{\rm d}\Omega}{4\pi}
\,\,\frac{v_\mu v^i E^i}{v\cdot D}\,,\eeq
involves a supplementary integration over the hard momenta $k$,
which, in terms of diagrams, corresponds
to closing the straight line in Fig.~\ref{D} into a hard loop.
Thus, the polarisation amplitudes generated by $j^{(0)}_\mu$
(cf. eq.~(\ref{exp})) are one-loop amplitudes
where the internal momentum is hard, while all the external 
lines are soft (or ultrasoft); some examples are shown in Fig.~\ref{C}.
These are precisely the {\it hard thermal loops}, which have
been originally computed from one-loop diagrams indeed \cite{BP90,FT90}.
As well known, the HTL's are only a part of the corresponding
one-loop amplitudes \cite{BP90,FT90,MLB96} (namely, the leading
order part for soft external lines), and this part is directly
singled out by the collisionless kinetic equation, 
cf. eqs.~(\ref{FL0}) and (\ref{jb0}) \cite{qcd,BIO96,prept}.

\begin{figure}
\protect \epsfxsize=14.5cm{\centerline{\epsfbox{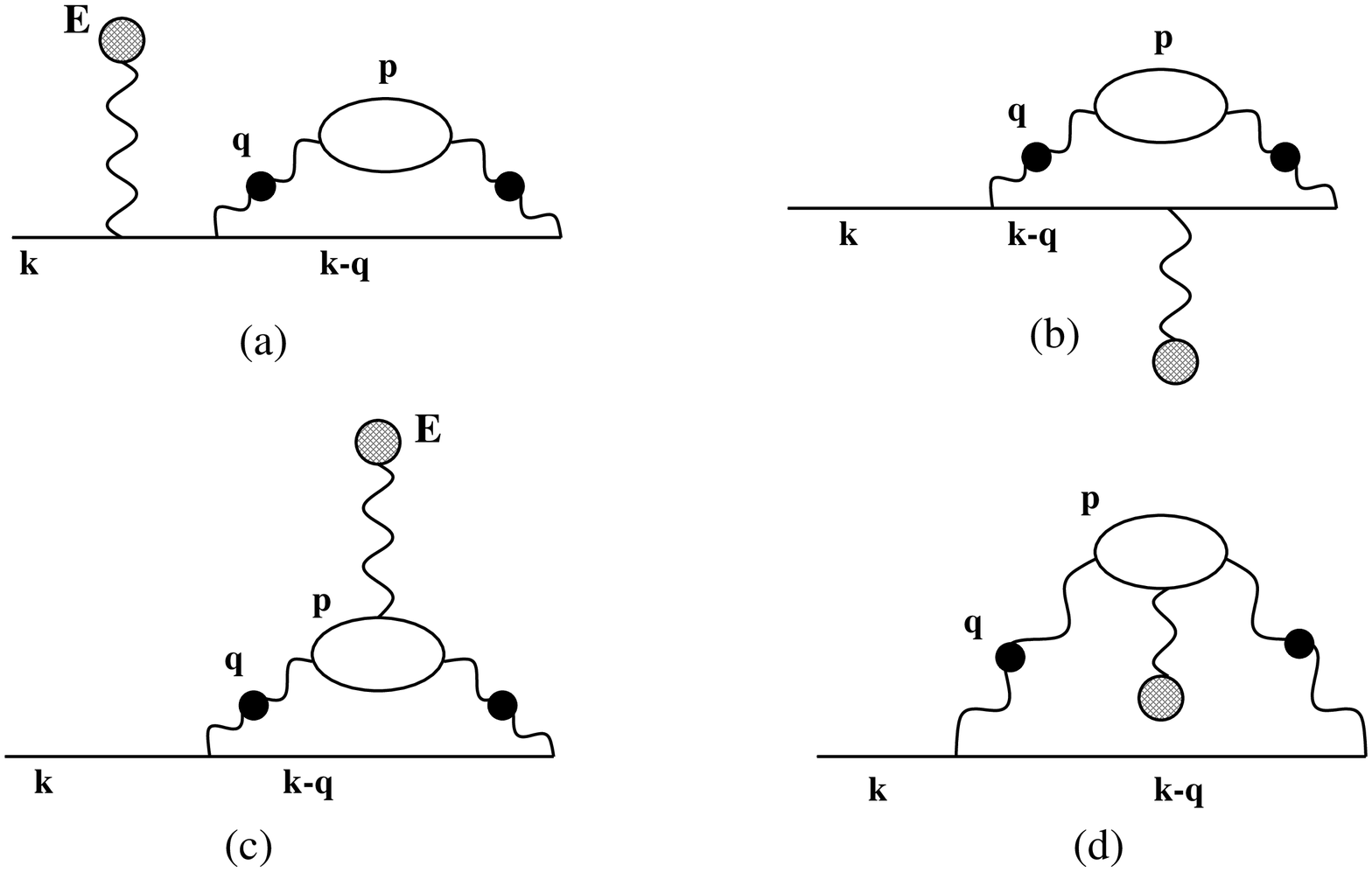}}}
         \caption{The first order iteration of the collision term
$C[\delta\acute G^{(0)}]$, to linear order in the background field.}
\label{E}
\end{figure}

\begin{figure}
\protect \epsfxsize=14.5cm{\centerline{\epsfbox{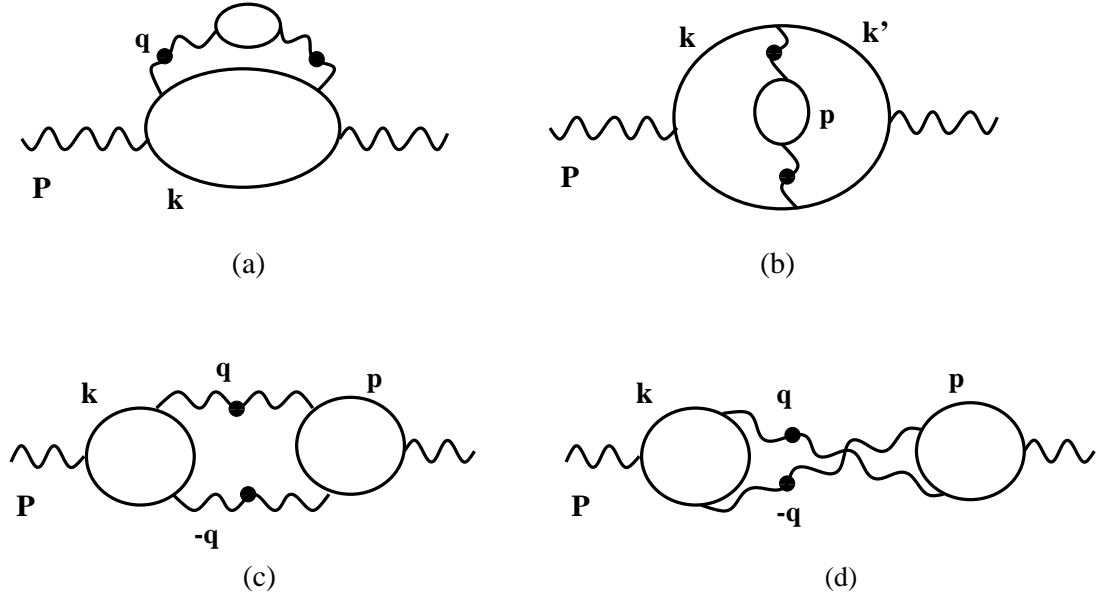}}}
         \caption{First order
 iteration for the ultrasoft polarization tensor.}
\label{A}
\end{figure}

Consider now the first order iteration of the collision term,
which yields:
\beq\label{FL1}
\delta \acute G^{(1)}\,=\,\frac{1}{v\cdot D}\,C[\delta
\acute G^{(0)}]\,.\eeq
The collision term $C[\delta 
\acute G^{(0)}]$ is illustrated in Fig.~\ref{E},
which should be compared to Fig.~\ref{FLS}.
For simplicity, we have only represented here single field insertions,
that is, we have linearized $C[\delta\acute G^{(0)}]$
 with respect to the
colour mean field. This is all what we need in order to
compute the first-order iteration of the ultrasoft
polarization tensor $\Pi_{\mu\nu}$.
The corresponding result is illustrated in Fig.~\ref{A},
and involves loop corrections to the corresponding HTL
(cf. Fig.~\ref{C}.a).

Before going on with higher iterations, let us make some comments
on the diagrams in Fig.~\ref{A}. These should be regarded as diagrams
of the thermal perturbation theory in {\it real time}, or linear
combinations of them. For instance, the self-energy insertion in
Fig.~\ref{A}.a stands for the combination $\Gamma(k)=\Sigma^<_{eq}(k)
 - \Sigma^>_{eq}(k)$ (cf. eq.~(\ref{EQG}))
and thus corresponds to the insertion of the quasiparticle damping rate
in the hard internal line. Similarly, the soft internal line
(with a bubble) in Fig.~\ref{A}.b stands for either
${}^*{\cal D}^<_{\mu\nu}(q)=-\Bigl({}^*{\cal D}_R
\Pi_{(0)}^<{}^*{\cal D}_A\Bigr)_{\mu\nu}$, or 
${}^*{\cal D}^>_{\mu\nu}(q)=-\Bigl({}^*{\cal D}_R
\Pi_{(0)}^>{}^*{\cal D}_A\Bigr)_{\mu\nu}$, where
$\Pi^{\mu\nu}_{(0)}$ denotes the two-point HTL
(cf. eq.~(\ref{HTL}) below), and
${}^*{\cal D}_{\mu\nu}$ is the HTL-resummed
propagator (the subscripts $R$ and $A$ refer, as usual, to
retarded and advanced propagators) \cite{BE}. To simplify the graphical
representation, it is convenient to replace
these graphs with the corresponding ones in the
{\it imaginary time} formalism, where the two diagrams in Figs.~\ref{A}.a
and b are replaced by the graphs in Fig.~\ref{PI2}.a and b, respectively,
while the diagrams in Figs.~\ref{A}.c and d remain formally the same,
and are globally represented in Fig.~\ref{PI2}.c. 
\begin{figure}
\protect \epsfxsize=13.5cm{\centerline{\epsfbox{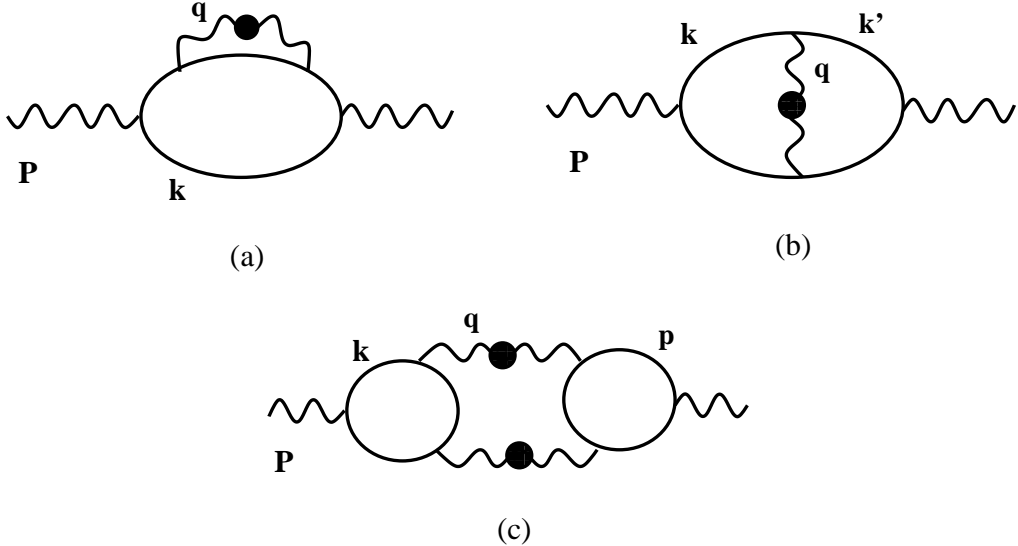}}}
         \caption{A simpler redrawing of the diagrams in Fig.~\ref{A}.}
\label{PI2}
\end{figure}

Now, since the Boltzmann equation has been obtained
from the exact field equations
by using various kinematical approximations \cite{BE},
the correspondence between its solution
(here, eq.~(\ref{FL1})) and the diagrams in Fig.~\ref{PI2} is only 
approximate: the kinetic equation isolates only the dominant
parts of these diagrams for ultrasoft external lines.
In fact, in the same way as the Vlasov equation (\ref{FL0})
provides the leading contribution to the one loop diagrams
when the external momenta are $\sim gT$, the Boltzmann equation
generates automatically the leading contributions to the
ultrasoft amplitudes when the external momenta are $\sim g^2 T$.
%This means that, if we start by computing the diagrams in
%in Fig.~\ref{PI2} (e.g., in the imaginary time formalism),
%and perform the same kinematical approximations as in the derivation
%the Boltzmann equation, then one should eventually obtain a result
%which is consistent with eq.~(\ref{FL1}) above.
Recently, this has been verified explicitly by B\"odeker
\cite{Bodeker99}, who computed the diagrams in Fig.~\ref{PI2}
and got the same result (namely, eq.~(\ref{PI1}) below)
as that obtained from  the iteration of the Boltzmann equation.
In fact, 
the approximations leading to the Boltzmann equation \cite{BE}
and those performed in Ref. \cite{Bodeker99} are similar. They all
rely upon the following chain of inequalities:
\beq\label{INSC}
\del_x\,\,\ll\,\,q\,\,\ll\,\, k\,,\eeq
which are controlled either by powers, or, at least, by a logarithm
of the coupling constant. Specifically:

({\it a}) The gauge covariant gradient expansion retains
the terms of leading order in $\del_x/k$;
this is an excellent approximation
since the neglected terms are of O$(g^2)$
or less. Diagramatically, this translates into the fact
that the smooth lines in Figs.~\ref{C}, \ref{A},
or \ref{PI2} (and also in the diagrams
to come) represent {\it eikonal} propagators, of the form
(cf. eq.~(\ref{GR})):
\beq\label{EIK0}
\Delta_{R}(P,{\bf v})\,=\,\frac{i}{v\cdot P + i\epsilon}
\,,\eeq
rather than standard tree-level propagators. Moreover, all the vertices
in these diagrams are simplified by systematically ignoring
the external momentum $P$. 

({\it b}) In the construction of the collision term,
we have retained only the terms of leading order in
an expansion in powers of $\del_x/q$. This approximation,
which assumes that both particles taking part in the collision 
(see Fig.~\ref{Born}) feel the same mean field, is needed in order 
to put the collision term into a form local in $x$.
But for colour fluctuations at the scale $g^2 T$, this is
correct only up to corrections of O$(1/\ln(1/g))\,$: indeed,
 $\del_x\sim g^2T$, while the cross section in eq.~(\ref{PHI})
is logarithmically sensitive to momenta $q\sim g^2 T$
(see eq.~(\ref{LLPHI}) below).
Diagrammatically, this affects only the diagram 
\ref{PI2}.c (more generally, the diagrams involving
two or more hard loops; see, e.g., Fig.~\ref{L} below), where it amounts
to assume that both the internal wavy lines (the two gluons connecting
two hard bubbles) carry the {\it same} momentum $q$.
(Strictly speaking, if one of these lines has a momentum ${\bf q}$,
then the other one should rather carry a momentum ${\bf p-q}$.)

({\it c}) Within the collision term, we have neglected, wherever 
possible, the exchanged momentum $q$ as compared to the hard momenta
of the colliding particles (recall the discussion after eq.~(\ref{CCOL})).
This is a good approximation since
$q/k ={\rm O}(g)$ or less. Diagrammatically, this entails more
simplifications in the propagators and vertices in 
Figs.~\ref{PI2}: the velocity remains unchanged when running along a
given hard loop (e.g., in Fig.~\ref{PI2}.c, there are only two velocities:
${\bf v}\equiv \hat{\bf k}$ for the left hand loop, and
${\bf v}^\prime \equiv \hat{\bf p}$ for the right hand one), 
and the momentum $q$ is neglected in all the vertices.

It is interesting to examine the validity of these approximations 
in the separate cases of colour fluctuations and
colourless ones. The approximations
({\it a}) and ({\it b}) are quite generic in relation with the 
Boltzmann equation, and are actually better justified for colourless 
fluctuations than for coloured ones: Indeed, we have seen in Sec. 1.1
that the colourless fluctuations relax mainly via hard (or large angle)
scattering, $q\sim T$, with a typical relaxation rate
$\sim g^4 T\ln(1/g)$. In this case,
the effect of the collisions becomes a leading order effect
only for very soft inhomogeneities, $\del_x \sim g^4 T$, for which
both inequalities $\del_x \ll k$ and $\del_x \ll q$ are very well
satisfied. On the other hand, the third approximation 
({\it c}) does not apply to colourless fluctuations, for which,
because of the cancellations discussed after eq.~(\ref{CLESS}),
$q\sim k\sim T$. Note that it is precisely approximation ({\it c})
which allowed us to reduce the original collision term (\ref{LINCOL})
to the simpler expression in eq.~(\ref{CSIMP}). 

The simplifications arising from approximation ({\it c})
(cf.  eqs.~(\ref{SIMP1}) and (\ref{SIMP2})) have actually a simple
diagrammatic interpretation as cancellations among Feynman graphs. 
The two diagrams in Figs.~\ref{PI2}.a and b correspond
respectively to the terms involving $W(k,x)$ and $W(k',x)$
in eq.~(\ref{LINCOL}). For colourless fluctuations in QCD, or,
equivalently, for electric fluctuations in QED, eq.~(\ref{CLESS})
teaches us that these two diagrams cancel each other in the
limit where $q$ is neglected next to $k$ or $p$. That is, each
of these diagrams is individually dominated by soft momenta $q$,
but their leading infrared contributions mutually cancel
in the sum of the diagrams, so that we are left with the (subleading)
contribution of hard $q$ momenta. (This is what makes the
colourless collision term (\ref{CLESS}) difficult to deal with;
see, e.g., \cite{BP90,Heisel94,Heisel94a}.) 

In QED, this cancellation has been also
verified via direct diagrammatic calculations, in Refs. 
\cite{Smilga90,Kraemmer94,Petit98,Bodeker99}.
In particular, in Ref. \cite{Bodeker99}, this has been related 
to the absence of HTL vertices with four external photons.
Indeed, the two diagrams in Figs.~\ref{PI2}.a and b
can be generated from the four-particle hard thermal loop in 
Fig.~\ref{C}.b, by closing two of the external lines in all the 
possible ways. In QED, this involves the four-photon HTL
which, however, is well known to vanish \cite{BP90,qcd} : the
HTL-like contributions of the individual diagrams with
four external photons (as in Fig.~\ref{C}.b) mutually cancel after 
summing over the permutations of the external lines.
In the present framework, the sum over the permutations
corresponds precisely to the sum of the two diagrams
in Figs.~\ref{PI2}.a and b, so this sum has to vanish as well.
%Similar arguments apply to the two diagrams in Figs.~\ref{A}.c and d,
%which in QED mutually cancel as well.

In QCD, on the other hand, the sum over permutations 
produces a colour commutator, which thus provides both a 
non-vanishing four-gluon HTL \cite{BP90,FT90}, and a non-zero
global contribution from the diagrams in Figs.~\ref{PI2}.
This is the content of eqs.~(\ref{CCOL})--(\ref{CSIMP}).
In fact, eq.~(\ref{SIMP1}) shows that, even in QCD, there remains a
{\it partial} compensation between the self-energy
and vertex corrections in Figs.~\ref{PI2}.a and b (while
the two diagrams in Fig.~\ref{A}.c and d rather reinforce each other;
cf. eq.~(\ref{SIMP2})). The net effect is that {\it half} of the
self-energy correction is cancelled by the vertex correction,
with the factor $1/2$ coming from the colour algebra (cf. the first
trace identity in eq.~(\ref{TRACE})).

\begin{figure}
\protect \epsfxsize=14.cm{\centerline{\epsfbox{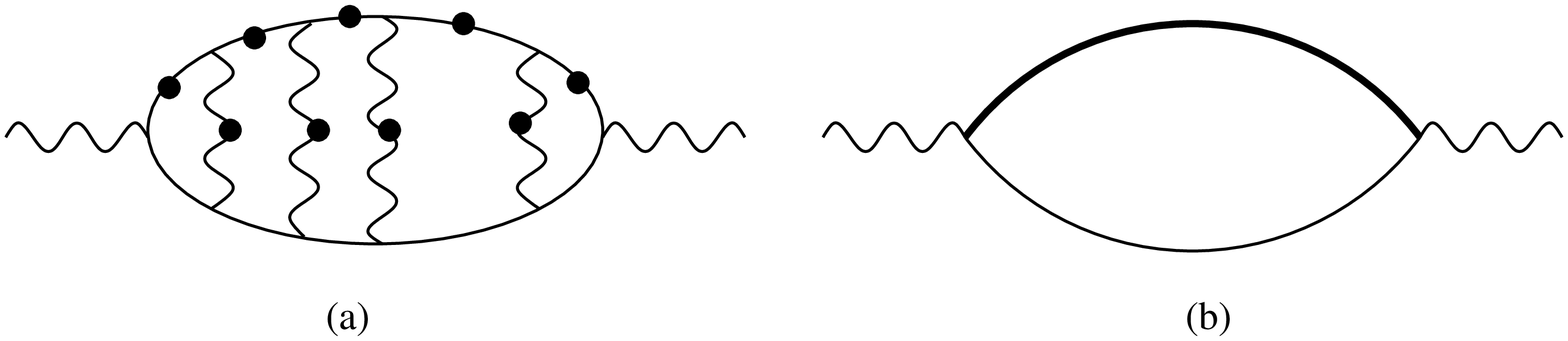}}}
         \caption{(a) A ladder diagram, as 
generated by iterations of the first
two pieces, $C_1$ and $C_2$, of the collision term; the smooth lines
with a bubble are eikonal propagators dressed with a damping rate $2\gamma$.
(b) The sum of all the ladders in (a),
as obtained after using the partial cancellation between
vertex and self-energy corrections to effectively remove the ladders;
the thick line is an eikonal propagator 
with a damping rate $\gamma$.}
\label{F}
\end{figure}
We now turn to the higher order iterations. Clearly, by iterating
the self-energy insertion in Fig.~\ref{FLS}.a, one ends up with replacing
the propagator of the hard gluon with a dressed propagator which includes the
damping rate. That is, the bare eikonal propagator
(\ref{EIK0}) is replaced by the following dressed propagator
\beq\label{DEIK0}
{}^*\Delta_{R}(P,{\bf v})\,=\,\frac{i}{v\cdot P + 2i\gamma}
\,,\eeq
to be graphically represented by a straight line with 
a blob (see Figs.~\ref{F}.a and ~\ref{L}). Equivalently, this resummation can
be achieved by moving the first collision term $C_1$ into
the l.h.s. of the Boltzmann equation (\ref{B}).
Similarly, by iterating the vertex correction in Fig.~\ref{FLS}.b 
one generates the ladders diagrams depicted in Fig.~\ref{F}.a. Finally,
diagrams with two or more hard loops will be generated by iterating the
other two pieces, $C_3$ and $C_4$, of the collision term
(cf. Figs.~\ref{FLS}.c and d, and Fig.~\ref{PI2}.c).
\begin{figure}
\protect \epsfxsize=14.cm{\centerline{\epsfbox{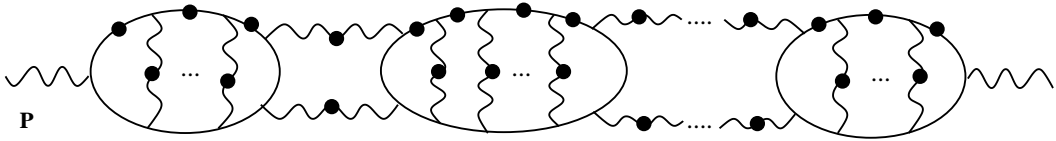}}}
         \caption{A generic ladder diagram contributing to the
ultrasoft polarization tensor, as obtained from the Boltzmann equation.}
\label{L}
\end{figure}

We conclude that the typical diagrams which are resummed
by the solution of the Boltzmann equation (\ref{B}) are 
as shown in Fig.~\ref{L}. They involve a chain of an arbitrary number
of hard loops, each of them dressed by ladders and damping effects
as in Fig.~\ref{F}.a, and connected one to the other by pairs of soft gluons.
The smooth lines with a blob represent the dressed eikonal
propagator (\ref{DEIK0}), while those without a blob are
thermal correlation functions like $G^>_0$ and $G^<_0$ 
(cf. eq.~(\ref{G0})), or derivatives of them (cf. eq.~(\ref{FL0})).
Now, all the previous examples
involve diagrams which contribute to the ultrasoft polarization
tensor (or 2-point amplitude) $\Pi_{\mu\nu}(P)$. But, of course,
similar diagrams exist for all the higher point ultrasoft
vertices: they can be obtained by inserting more external
lines along the hard loops in Fig.~\ref{L}, on any of the
internal eikonal lines (which, in general, can be seen
as eikonal propagators in a background field; cf. eq.~(\ref{GR})).

It is finally possible to give a simple graphical interpretation of the
(partial) infrared cancellations between self-energy and vertex
corrections, as discussed above. To the order of interest, the only
effect of the ladder corrections in  Fig.~\ref{F}.a is to reduce to damping
rate\footnote{In QED, the sum of all the self-energy and 
ladder corrections depicted in Fig.~\ref{F}.a simply vanishes
to the order of interest \cite{Smilga90}; cf. the discussion
after eq.~(\ref{CLESS}).}
in eikonal propagators like (\ref{DEIK0}) from
$2\gamma$ to $\gamma$ (cf. eq.~(\ref{SIMP1})). This is depicted
in Fig.~\ref{F}.b, where the thick internal line denotes the
following eikonal propagator (compare to eqs.~(\ref{EIK0})
and (\ref{DEIK0})):
\beq\label{AEIK0}
\acute\Delta_{R}(P,{\bf v})\,=\,\frac{i}{v\cdot P + i\gamma}
\,\eeq
while the thin line corresponds to $\del_k G^<_0(k)$.

\subsection{General properties and iterative solutions}

We now return to a discussion of the general properties
of the ultrasoft amplitudes. Since they encompass, and
generalize, the HTL's, we expect these amplitudes to describe
phenomena like Debye screening or Landau damping
(possibly modified by the effects of the collisions), and also
transport phenomena, which are made possible by the collision term;
the exemple of the colour conductivity will be discussed in the
next section.

In order to look for Debye screening, it is enough to consider static
fields, that is, colour field configurations which are
described by time-independent vector potentials $A^\mu_a ({\bf x})$.
In this case, the ultrasoft amplitudes reduce to the
usual Debye mass term $m_D^2= g^2 NT^2/3$ for electrostatic fields,
as obtained from the HTL's.
In order to see this, it is convenient to
decompose the functions $W^a(x,{\bf v})$ as follows \cite{qcd} :
\beq\label{CALA}
W^a(x,{\bf v})\,\equiv\,-A_0^a(x)\,+\,{\cal A}^a(x,{\bf v}).\eeq
From qs.~(\ref{W10}) and (\ref{CALA}), the following equation
is obtained (recall that $E^i_a=D^i_{ab}A^0_b-\del^0A^i_a$) :
\beq\label{W11}
(v\cdot D_x)^{ab}{\cal A}_b(x,{\bf v})&=&\del^0(v\cdot A^a)
\,-\,\gamma\Bigl\{{\cal A}^a(x,{\bf v})\,-\,\langle 
{\cal A}^a(x,{\bf v})\rangle \Bigr\},\eeq
while eq.~(\ref{j1}) shows that the current can be rewritten as:
\beq\label{j2}
j^\mu_a(x)\,=\,-\,\delta^{\mu 0}\,m^2_{D} A_0^a(x)\,+\,
m_D^2\int\frac{{\rm d}\Omega}{4\pi}
\,v^\mu{\cal A}_a(x,{\bf v}).\eeq
In obtaining eq.~(\ref{W11}), we have also used the fact that:
\beq\label{NULL}
A_a^0(x)\,-\,\langle A^0_a(x)\rangle\,=\,0\,,\eeq
since the collision term vanishes for any function which is
independent of ${\bf v}$.

In eq.~(\ref{W11}), the time derivative of
the vector potentials (i.e., the term $\del^0(v\cdot A)$)
acts as a source for the functions ${\cal A}^a(x,{\bf v})$.
Since we are looking here for solutions which vanish in the absence
of sources, it follows that ${\cal A}^a(x,{\bf v})=0$
(and therefore  $W^a({\bf x},{\bf v})=-A_0^a ({\bf x})$)
when the gauge potentials are time-independent.
Then, eq.~(\ref{j2}) reduces to:
\beq\label{jstat}
j_\mu^{a}({\bf x})
\,=\,-\,\delta_{\mu 0}\,m^2_{D} A_0^a({\bf x})\,,\eeq
which is the same expression as in the HTL approximation \cite{qcd}. 
That is, for static external legs, all the ultrasoft vertices
with $n\ge 3$ external lines vanish, while $\Pi_{\mu\nu}
(\omega=0,{\bf p})=-\delta_{\mu 0}\delta_{\nu 0}\,m_D^2$.

Eq.~(\ref{jstat}) shows, in particular, that the value of the
Debye mass is not modified by the collisions among the hard
particles. An alternative derivation of this result has been
recently given in Ref. \cite{ASY99} (see also Sec. 4.2 below).
This is not unexpected since we know
\cite{Rebhan93} that the first correction to $m_D^2$,
of O$(g^3T^2\ln(1/g))$, comes out from the interactions between
soft and ultrasoft fields. 

For time-dependent fields, however,
the collisions among the hard particles do play a role,
and, at very soft momenta $P\simle g^2 T$ (by which we mean that
both the frequency $P^0\equiv \omega$, and the spatial
momentum $p=|{\bf p}|$, are of order $g^2 T$ or less), they
can even dominate over the mean field effects.
This may be seen by considering the formal solution
of the Boltzmann equation (\ref{W10}) obtained by iterations.
There are several ways to organize the iteration.
For instance, we may iterate the whole collison term in the
r.h.s. of eq.~(\ref{W10}), similarly to what we have done
in the previous subsection.
Since the collision term is proportional to $\gamma$, the resulting
solution is a formal expansion in powers of $\gamma$.
Specifically, we write $W=W^{(0)}+ W^{(1)}+W^{(2)}+\,\dots$, where
$W_a^{(0)}(x,{\bf v})$ satisfies the transport equation in the mean
field approximation (or Vlasov equation)
\beq\label{WVL}
(v\cdot D_x)^{ab}W_b^{(0)}(x,{\bf v})&=&{\bf v}\cdot{\bf E}^a(x),\eeq
while the $N$th order correction $W_a^{(N)}(x,{\bf v})$
is proportional to $\gamma^N$.
The (retarded) solution to eq.~(\ref{WVL}) involves the eikonal
propagator $\Delta_R(x,y;{\bf v})$, as defined in eq.~(\ref{GR}).
Thus, in compact notations:
\beq\label{ITER0}
W^{(0)}&=&\frac{{\bf v}\cdot{\bf E}}{v\cdot D}\,,
\nonumber\\
W^{(1)}&=&-\,\frac{\gamma}{v\cdot D}\left\{
\frac{{\bf v}\cdot{\bf E}}{v\cdot D}\,-\,\left\langle
\frac{{\bf v}\cdot{\bf E}}{v\cdot D}\right\rangle\right\},
\nonumber\\
W^{(N)}&=&-\,\frac{\gamma}{v\cdot D}
\Bigl\{W^{(N-1)}\,-\,\Bigl\langle W^{(N-1)}\Bigr\rangle\Bigr\}.\eeq
This expansion maintains explicit gauge symmetry at each
order in $\gamma$: indeed, since both pieces of the collision term 
(i.e. the local piece $-\gamma W^a(x,{\bf v})$ and the non-local
one $\gamma \langle W^a(x,{\bf v})\rangle$) are treated on the same
footing, the current conservation law (\ref{COVCONS}) is verified
at each step in this iteration. Then, e.g., the polarization 
tensor $\Pi_{\mu\nu}^{(N)}$ constructed in the $N$th iteration is
guaranteed to be transverse.

The discussion in the previous subsection provides us with
the diagramatic interpretation of the expansion (\ref{ITER0}).
The zeroth order solution $W^{(0)}$ corresponds obviously
to the mean field, or HTL, approximation 
(cf. eqs.~(\ref{FL0}) and (\ref{jb0}), and Figs.~\ref{D} and \ref{C}).
The first order iteration $W^{(1)}$ corresponds to the diagrams
in Figs.~\ref{A} or \ref{PI2}. Specifically, the expression of
$W^{(1)}$ in eq.~(\ref{ITER0}) involves two pieces within the braces
in its r.h.s.: the first piece corresponds to the {\it sum}
of the self-energy and vertex corrections depicted
in Figs.~\ref{A}.a and b (or, equivalently, in 
Figs.~\ref{PI2}.a and b);
similarly, the second piece in $W^{(1)}$ corresponds to the sum of
the two diagrams in Figs.~\ref{A}.c and d. But it is only the set
of the four diagrams in Fig.~\ref{A} which is globally gauge
invariant and provides a transverse contribution $\Pi_{\mu\nu}^{(1)}$
to the polarization tensor \cite{Bodeker99} 
(see also Sec. 4.1 below, especially eq.~(\ref{PI1})). 
 Note that, for ultrasoft gradients
$D_x\sim g^2 T \sim \gamma$, the expansion (\ref{ITER0})
is a formal one: indeed, all the terms are equally important.

A different expansion is obtained by choosing only the
second piece of the collision term, namely
$\gamma\langle W^a(x,{\bf v})\rangle$, as the perturbation.
To do this, we move the term $-\gamma W^a(x,{\bf v})$
to the l.h.s. of the Boltzmann equation, and
define the ``dressed'' eikonal propagator:
\beq\label{GRG}
\acute\Delta_{R}^{ab}(x,y;{\bf v})\,=\,\theta (t)\,\delta^{(3)}
\left({{\bf x}}-{{\bf y}}-{\bf v}t
\right )\,{\rm e}^{-\gamma t}\,U^{ab}(x,y)
\,\equiv\,\Bigl\langle x,\,a\Bigl|\frac{i}
{i(v\cdot D)+i\gamma}\Bigr|y,\,b\Bigr\rangle.\eeq
This operation looks naively like a resummation of the damping
rate $\gamma$
in the propagator of the hard particles, but in reality it is the
combined effect of a self-energy and a vertex resummation
(recall the discussion after 
eq.~(\ref{CCOL}), and also the diagrams in 
Figs.~\ref{F}.a and b);
the resummation of the self-energy alone would have
given an attenuation factor $2\gamma$.
The corresponding iterative solution reads then:
\beq\label{ITER1}
W^{(0)}&=&\frac{{\bf v}\cdot{\bf E}}{v\cdot D+\gamma}\,,
\nonumber\\
W^{(1)}&=&\frac{\gamma}{v\cdot D+\gamma}\left\langle
\frac{{\bf v}\cdot{\bf E}}{v\cdot D+\gamma}\right\rangle,
\nonumber\\
W^{(N)}&=&\frac{\gamma}{v\cdot D+\gamma}
\Bigl\langle W^{(N-1)}\Bigr\rangle,\eeq
and suggests that, for very soft colour mean fields, the damping 
rate $\gamma\sim g^2 T\ln(1/g)$ may act as an effective IR cutoff.
That is, for gradients $D_x\simle g^2 T$, we can even neglect the drift term
$v\cdot D$ as compared to $\gamma$ (at least, to leading logarithmic 
accuracy; cf. Sec. 3.4 below), in which case the expansion
(\ref{ITER1}) may be resummed into an exact solution (cf. Sec. 4.2).

Diagramatically,  the zeroth order solution $W^{(0)}$ in 
eq.~(\ref{ITER1}) corresponds to Fig.~\ref{F}.b, i.e., to the sum
of all the ladder diagrams in Fig.~\ref{F}.
(Incidentally, this is also equivalent to the relaxation time
approximation, eq.~(\ref{RTA}).)
This is not a gauge-invariant subset of diagrams, and, indeed,
it is quite obvious that
the expansion eq.~(\ref{ITER1}) violates gauge symmetry at 
any finite order (since it treats the two pieces of the collision term
on a different footing); see also the discussion at the end
of Sec. 4.1.

\subsection{The leading-logarithmic approximation}

The previous applications of the Boltzmann equation (\ref{W1})
\cite{Bodeker98,ASY98,ASY99,Bodeker99}
have been mostly limited to the leading-logarithmic approximation
(LLA) that we shall describe now.

 Recall first that, for colour
excitations at the scale $g^2 T$, the collision term in
eq.~(\ref{W1}) is known, strictly speaking, only to logarithmic
accuracy, that is, up to corrections of O$(1/\ln(1/g))\,$.
This limitation has two sources: {\it i}) the IR problem
of the damping rate \cite{BP90,Smilga90,Pisarski93,lifetime},
and {\it ii}) the gradient expansion in the presence of
long range interactions \cite{BE} (i.e., the expansion in
powers of $\del_x/q$; cf. Sec. 3.2). 
Because of that, it has been previously argued that the
Boltzmann equation should be further simplified, for consistency,
so as to preserve only the terms which are enhanced by a
logarithm. Specifically, this involves two approximations:

{\it a}) In eq.~(\ref{PHII}) for the collision integral
$\Phi({\bf v\cdot v}^\prime)$ one has retained only the singular
piece of the magnetic propagator, namely \cite{lifetime}:
\beq \label{singDT}
|{}^*{\cal D}_t(q_0 \ll q)|^2\,\simeq\,
\frac{1} {q^4 + (\pi m_D^2 q_0/4q)^2}\,\,\,
\longrightarrow_{q\to 0}\,\,\,\frac{4}{m_D^2}\,\frac{\delta(q_0)}{q}\,.\eeq
This allows one to isolate the IR singular piece of eq.~(\ref{PHII}),
which reads \cite{Bodeker98} :
\beq\label{LLPHI}
\Phi({\bf v\cdot v}^\prime)\,\simeq\,\Phi_0({\bf v\cdot v}^\prime)
\,\equiv\,\frac{2}{\pi^2 m_D^2}\,
\frac{({\bf v\cdot v}^\prime)^2}{\sqrt{1-({\bf v\cdot v}^\prime)^2}}\,
\ln\,\frac{1}{g}\,,
\eeq
where the logarithm $\ln(1/g)$ in the r.h.s. 
has been generated via the following integral:
\beq\label{LLES}
\int_{\mu}^{m_D}\frac{{\rm d}q}{q}\,=\,\ln\frac{m_D}{\mu}\,\simeq\,
\ln\,\frac{1}{g}\,.\eeq
In this equation, the upper cutoff $m_D$ is given by the screening effects
at the scale $gT$ (as included in ${}^*{\cal D}_t$, eq.~(\ref{singDT})),
while the IR cutoff $\mu$ is either the non-perturbative ``magnetic
mass''  \cite{MLB96} (in which case $\mu\sim g^2 T$), or --- in the
framework of the effective theory for ultrasoft
fields  \cite{Bodeker98} ---,
the intermediate scale $\mu\simeq g^2T\ln(1/g)$ separating ultrasoft from
soft momenta. In both cases, the estimate (\ref{LLES})
holds to leading-log accuracy.
By inserting the approximation (\ref{LLPHI}) into eq.~(\ref{gamma1}),
we get the damping rate to the same accuracy ($\alpha =g^2/4\pi$) :
\beq\label{gamma3}
\gamma\,\simeq\,\gamma_0\,\equiv\,\alpha NT\,\ln\frac{1}{g}\,.\eeq
In fact, the 
expression of $\gamma$ obtained by evaluating exactly the
integrals in eqs.~(\ref{PHII}) and (\ref{gamma1}) with a sharp
IR momentum cutoff  $\mu$ (see Appendix B in the last paper
of Ref. \cite{lifetime}) is:
\beq\label{gammamu}
\gamma\,=\,\alpha NT\,\ln\frac{m_D}{\mu}\,,\eeq
up to corrections of order\footnote{Actually,
numerical studies of eqs.~(\ref{PHII}) and (\ref{gamma1}) 
show that the error is even smaller, of order $(\mu/m_D)^2$
\cite{tony}.} $\mu/m_D$.

{\it b}) The covariant gradient operator, or drift term,
$v\cdot D_x\sim g^2 T$ in the l.h.s. of eq.~(\ref{W1}) has been neglected
next to the collision term $\propto \gamma$ in the r.h.s.

After these simplifications, eq.~(\ref{W1}) reduces to:
\beq\label{W2}
{\bf v}\cdot{\bf E}^a(x)\,=\,\gamma_0
\Bigl\{W^a(x,{\bf v})\,-\,\langle W^a(x,{\bf v})\rangle_0
\Bigr\},\eeq
where the subscript 0 refers to the LLA,
cf. eqs.~(\ref{LLPHI}) and (\ref{gamma3})
(e.g., the angular average $\langle W^a(x,{\bf v})\rangle_0$
is given by eq.~(\ref{AVF}) with $\Phi\to \Phi_0$).

Eq.~(\ref{W2}) can be easily solved by iterations, as in
eq.~(\ref{ITER1}): the first iteration yields $W^{(0)}=
{\bf v}\cdot{\bf E}/\gamma_0$, and all the higher order iterations
vanish ($W^{(N)}=0$ for $N\ge 1$) since  $W^{(0)}$ is an
odd function of ${\bf v}$, while $\Phi_0({\bf v\cdot v}^\prime)$
is even (cf. eq.~(\ref{LLPHI})). Thus,
\beq\label{LLW}
W^a({\bf v})\,\simeq\,\frac{{\bf v}\cdot {\bf E}^a}{\gamma_0}\,,
\eeq 
which, as already mentioned in the Introduction, is formally equivalent
to the relaxation time approximation (\ref{RTA}), and generates a
colour current ${\bf j}^a =\sigma_0{\bf E}^a\,$,
with the colour conductivity in the LLA
$\sigma_0=m_D^2/3\gamma_0$ \cite{Bodeker98,ASY98}.

However, the approximation ({\it b}) above is insufficient for several
reasons:\\ {\it i}) It is incorrect in the electric sector, where it fails
to provide Debye screening \cite{ASY99}. 
Indeed, eq.~(\ref{LLW}) yields $j_a^0=0$, 
to be contrasted with the correct result (\ref{jstat}):
$j^0_{a}=-m^2_{D} A_0^a({\bf x})$. This is so since, for static fields,
 $W^a({\bf x},{\bf v})=-A_0^a ({\bf x})$ is an exact solution of
eq.~(\ref{W1}), for which the collision term vanishes
(cf. eq.~(\ref{NULL})); in this case,
it  is not legitimate to neglect the drift term.\\
{\it ii}) In some specific kinematical situations (essentially, for
fields which are arbitrarily weak and slowly varying), the 
linearized  Boltzmann equation can be solved  to a higher accuracy than in
 the LLA, leading to a formula for  the transverse colour conductivity
valid beyond the LLA. This will be explained in Sec. 4.2 below.\\
{\it iii}) In order to study the non-local
structure of the ultrasoft amplitudes, one has
to retain the drift term in the Boltzmann equation. 
Then, both pieces of the collision term (local or non-local
in ${\bf v}$, cf. eq.~(\ref{W1})) play a role,
as required by gauge symmetry (cf. eq.~(\ref{COVCONS})).
This will be further explained on the example of the
polarization tensor, in the next section.

\setcounter{equation}{0}
\section{The polarization tensor}

In this section we shall discuss in more detail the polarization
tensor $\Pi_{\mu\nu}(P)$ for ultrasoft ($P\simle g^2T$) fields,
as determined by the solution to the Boltzmann equation.
The  polarization tensor typifies the
non-local structure of all the ultrasoft amplitudes: as in the HTL
approximation,  all the $n$-point vertices
with $n\ge 3$ follow from it as a consequence of the non-Abelian gauge symmetry
(these vertices are generated by the covariant derivative in
the l.h.s. of the Boltzmann equation (\ref{W1})).

\subsection{Tensorial structure and iterative solutions}

According to eq.~(\ref{exp}), in order to construct the polarization
tensor it is enough to consider a linearized
version of the Boltzmann equation (\ref{W10}), namely:
\beq\label{WLIN}
(v\cdot \del_x)W^a(x,{\bf v})&=&{\bf v}\cdot{\bf E}^a(x)-
\gamma\Bigl\{W^a(x,{\bf v})\,-\,\langle W^a(x,{\bf v})\rangle
\Bigr\},\eeq
where $E^i_a(x)\equiv \del^iA^0_a-\del^0A^i_a$ denotes only the
``Abelian'' piece of the electric mean field.
It is then useful to go to the
momentum representation, where eq.~(\ref{WLIN}) becomes
\beq\label{LINI}
(v\cdot P)W(P,{\bf v})&=&i{\bf v}\cdot{\bf E}(P)\,-\,
i\gamma\Bigl\{W(P,{\bf v})\,-\,\langle W(P,{\bf v})\rangle
\Bigr\},\eeq
with $P^\mu=(\omega,{\bf p})$, $E^i(P)= i(\omega A^i(P)-p^iA^0(P))$,
and the colour indices have been omitted since trivial: the linearized
equations (\ref{WLIN}) or (\ref{LINI}) are indeed diagonal in colour.

Even though linear, eq.~(\ref{LINI}) is still difficult to solve in general,
since the term $\gamma\langle W^a(x,{\bf v})\rangle$
is non-local in ${\bf v}$ (cf. eq.~(\ref{AVF})).
Below, we shall consider an iterative solution, following the 
procedure explained at the end of Sec. 3.3. But before doing that,
we shall derive from eq.~(\ref{LINI})
some general properties of $\Pi_{\mu\nu}(P)$.

First, the solution $W(P,{\bf v})$ can be written in the form:
\beq\label{WI}
W(P,{\bf v})&=&i\,W^i(P,{\bf v})E^i(P),\eeq
with the new functions $W^i(P,{\bf v})$ satisfying:
\beq\label{LIN}
(v\cdot P)W^i(P,{\bf v})&=&v^i\,-\,
i\gamma\Bigl\{W^i(P,{\bf v})\,-\,\langle W^i(P,{\bf v})\rangle
\Bigr\}.\eeq
The corresponding colour current can then be written as:
\beq\label{jlin}
j^\mu(P)\,\equiv\, m_D^2\int\frac{{\rm d}\Omega}{4\pi}
\,v^\mu\,W(P,{\bf v})\,=\,\sigma^{\mu i}(P)E^i(P),\eeq
with the following conductivity tensor:
\beq\label{DEFSIG}
\sigma^{\mu i}(P)\,\equiv\,im_D^2\int\frac{{\rm d}\Omega}{4\pi}
\,v^\mu\,W^i(P,{\bf v}).\eeq
The polarization tensor is then defined by
$j^\mu (P)\equiv \Pi^{\mu\nu}(P)A_\nu(P)$, which implies:
\beq\label{PIS}
\Pi^{\mu 0}(P)\,=\,-ip^j\sigma^{\mu j}(P),\qquad
\Pi^{\mu i}(P)\,=\,-i\omega\sigma^{\mu i}(P).\eeq
By using eq.~(\ref{LIN}), one can verify that 
the resulting polarization tensor is transverse,
$P_\mu\Pi^{\mu\nu}=0$, and symmetric, $\Pi^{\mu\nu}=\Pi^{\nu\mu}$.
These properties, however, are not manifest on eqs.~(\ref{DEFSIG})
and (\ref{PIS}).

The transversality property reflects the conservation of
the (linearized) current ($P_\mu j^\mu=0$) and has been already proven
in a more general context in Sec. 3.1 (recall eq.~(\ref{Wtr})).
Here, it immediately follows from eq.~(\ref{LIN}), which
implies (compare to eq.~(\ref{VERIF})):
\beq\label{CWL}
\int\frac{{\rm d}\Omega}{4\pi}\,(v\cdot P)W^i(P,{\bf v})&=&0,\eeq
so that $P_\mu\sigma^{\mu i}=0$, and hence 
$P_\mu\Pi^{\mu\nu}=0$. For what follows, it is useful to 
decompose $W^i(P,{\bf v})$ into its
longitudinal and transverse components
with respect to ${\bf p}$, by writing (with $\hat p^i = p^i/p$
and $p = |{\bf p}|$) :
\beq\label{WL}
W^i(P,{\bf v})\,=\,W_L(P,{\bf v})\hat p^i + W^i_T(P,{\bf v}),
\qquad W_L\equiv {\hat{\bf p} \cdot{\bf W}},\qquad
{\bf p\cdot  W}_T=0,\eeq
and note that
eq.~(\ref{CWL}) entails a constraint on the longitudinal 
component alone:
\beq\label{CWL1}
\int\frac{{\rm d}\Omega}{4\pi}\,(\omega-{\bf v\cdot p})
W_L(P,{\bf v})&=&0.\eeq
By using this constraint, together with the properties
of the angular integration, we shall now verify
the symmetry property $\Pi^{\mu\nu}=\Pi^{\nu\mu}$.
Eqs.~(\ref{DEFSIG}) and (\ref{PIS}) imply, e.g., 
\beq
\Pi^{0i}(P)\,=\,\omega
m_D^2\int\frac{{\rm d}\Omega}{4\pi}\,W^i(P,{\bf v}),\qquad
\Pi^{i0}(P)\,=\,p m_D^2\int\frac{{\rm d}\Omega}{4\pi}\,v^i\,
W_L(P,{\bf v}).\eeq These two expressions are indeed identical
because:
\beq
\Pi^{i0}(P)\,=\,\hat p^i\,m_D^2\int\frac{{\rm d}\Omega}{4\pi}\,
({\bf v}\cdot{\bf p})W_L
\,=\,\hat p^i\,\omega
m_D^2\int\frac{{\rm d}\Omega}{4\pi}\,W_L\,=\,\Pi^{0i}(P).\eeq
In writing the first equality above, we have used the fact that
$\hat{\bf p}$ is the only remaining vector after performing 
the integral over ${\bf v}$; then, eq.~(\ref{CWL1}) has been
used to obtain the second equality. It can be similarly shown that
$\Pi^{ij}=\Pi^{ji}$.

The above properties fix the tensor structure of $\Pi^{\mu\nu}$:
as in the HTL approximation, $\Pi^{\mu\nu}$ is determined by two independent
scalar functions $\Pi_L(\omega,p)$ and $\Pi_T(\omega, p)$, which we 
choose as:
\beq\label{PILT}
\Pi_L(\omega,p)&\equiv&-\Pi^{00}(P)\,=\,
-p m_D^2\int\frac{{\rm d}\Omega}{4\pi}\,W_L(P,{\bf v}),\nonumber\\
\Pi_T(\omega, p)&\equiv&\frac{1}{2}(\delta_{ij}-\hat p_i\hat p_j)
\Pi^{ij}(P)\,=\,\frac{1}{2}\,\omega
m_D^2\int\frac{{\rm d}\Omega}{4\pi}\,{\bf v\cdot W}_T(P,{\bf v}).\eeq
In terms of these functions, the components of  $\Pi^{\mu\nu}$ read:
 \beq\label{Pitens}
\Pi^{00}(P)= -\Pi_L(\omega,{p}),\qquad
\Pi^{0i}(p)= -\,\frac{\omega p^i}{p^2}\,
 \Pi_L(\omega,{p}),\\ \nonumber
\Pi^{ij}(P)=(\delta^{ij}-\hat p^i\hat p^j) \Pi_T(\omega,{p})
- \hat p^i\hat p^j \,\frac{\omega^2}{p^2}\,\Pi_L(\omega,{p})\,.\eeq
As $p\to 0$, there is no privileged direction, and, since
$\Pi_T(\omega,p=0)$ is non-zero (e.g., $\Pi_T(\omega,p=0)
= m_D^2/3 \equiv \omega_{pl}^2$ in the HTL approximation
\cite{MLB96}), the above expression for $\Pi^{ij}$ requires
$\Pi_L(\omega,p\to 0)$ to vanish in the following
way:
\beq\label{PL0}
\Pi_L(\omega,p\to 0)\,\approx\,-\,\frac{p^2}{\omega^2}\, \Pi_T(\omega,p=0).
\eeq
The longitudinal and transverse components
of the conductivity tensor will be also needed later. Writing
$j^i=\hat p^i j_L+j_T^i$ and $E^i=\hat p^i E_L + E_T^i$
(with $E_L=i(\omega A_L - p A_0)$
and $E_T^i=i\omega A_T^i$), and defining $\sigma_L$ and $\sigma_T$
such that $j_L=\sigma_L E_L$ and $j_T^i=\sigma_T E_T^i$, we get:
\beq\label{SLT}
\sigma_L(\omega,p)\,=\,-i\,\frac{\omega}{p^2}\,\Pi_L(\omega,p),
\qquad\sigma_T(\omega,p)\,=\,\frac{i}{\omega}\,\Pi_T(\omega,p).
\eeq

We now turn to a discussion of the iterative solution to
eq.~(\ref{LIN}), following the considerations at the end of Sec. 3.3.
 If we treat the whole
collision term as a perturbation, then the first two iterations read
(cf. eq.~(\ref{ITER0})):
\beq\label{ITER2}
W_i^{(0)}\,=\,\frac{v_i}{v\cdot P}\,,\qquad
W_i^{(1)}\,=\,-i\,\frac{\gamma}{v\cdot P}\left\{
\frac{v_i}{v\cdot P}\,-\,\left\langle
\frac{v_i}{v\cdot P}\right\rangle\right\}.\eeq
The first term above yields the well known HTL approximation for
the (retarded) polarization tensor \cite{BIO96,MLB96}, namely:
\beq\label{HTL} \Pi_{\mu\nu}^{(0)}(\omega, {\bf  p}) \,=\,
m_D^2
\left \{-\delta^0_\mu\delta^0_\nu \,+\,\omega \int\frac{{\rm d}\Omega}{4\pi}
\,\frac{v_\mu\, v_\nu} {\omega - {\bf  v}\cdot {\bf  p}
+i\epsilon}\right\}\,.\eeq
The second term in eq.~(\ref{ITER2})
gives then a correction to the HTL result which can be 
written in the following form:
\beq\label{PI1} \Pi_{\mu\nu}^{(1)}(P)&=&-i\gamma \omega m_D^2
\int\frac{{\rm d}\Omega}{4\pi}\,
\frac{v_\mu} {v\cdot P}\left\{
\frac{v_\nu}{v\cdot P}\,-\,\left\langle
\frac{v_\nu}{v\cdot P}\right\rangle\right\}\nonumber\\&=&
-i \omega m_D^4\,\frac{g^2 N T}{2}\int\frac{{\rm d}\Omega}{4\pi}
\int\frac{{\rm d}\Omega'}{4\pi}
\,\Phi({\bf v\cdot v}^\prime)\,
\frac{v_\mu} {v\cdot P}\left\{
\frac{v_\nu}{v\cdot P}\,-\,
\frac{v^{\prime}_\nu}{v'\cdot P}\right\},\eeq
where in the second line we have used the definition
(\ref{AVF}) of the angular averaging together with eq.~(\ref{gamma1})
for $\gamma$. Higher order iterations $\Pi_{\mu\nu}^{(N)}$
can be written down similarly, in a straightforward way.

With the leading logarithmic approximation
(\ref{LLPHI}) for $\Phi({\bf v\cdot v}^\prime)$, 
eq.~(\ref{PI1}) coincides with the expression recently obtained 
by B\"odeker in
Ref. \cite{Bodeker99} by diagrammatic calculations.
From the discussion in Sec. 3.2, one easily associates
the first term within the braces in eq.~(\ref{PI1}) with
the two diagrams in Figs.~\ref{PI2}.a 
and b, and the second term, which is
non-local in ${\bf v}$, to the diagram with two
hard loops in Fig.~\ref{PI2}.c (the two unit vectors
${\bf v}$ and ${\bf v}'$ correspond to the velocities of the hard
particles running around these
two loops). The higher-order
iterations $\Pi_{\mu\nu}^{(N)}$ with $N\ge 2$ would similarly
correspond to the diagrams illustrated in Fig.~\ref{L}.

For $P\sim g^2 T$,
however, the contribution in eq.~(\ref{PI1}) is actually of 
the same order in $g$ as the HTL (\ref{HTL}), and even dominates over
the latter by a logarithm $\ln(1/g)$. This reflects the fact,
already emphasized in Sec. 3.3, that the iterative expansion is
generally not appropriate for the problem at hand, and
it makes a priori no sense to try and evaluate $\Pi_{\mu\nu}$ from
just a finite number of terms in this expansion.
In the next subsection, we shall rather construct exact solutions
to the equation (\ref{LIN}) in specific kinematical limits.

Consider finally the second iterative solution, as described in
eq.~(\ref{ITER1}). In Sec. 3.4, this proved to be useful in
obtaining the leading-logarithmic estimate in eq.~(\ref{LLW}).
In general, however, this expansion must be used with caution since,
as advertised at the end of Sec. 3.3, it violates gauge symmetry at 
every finite order. For instance, if we restrict ourself to the
zeroth order iteration,
we obtain (cf. eqs.~(\ref{LIN}) and (\ref{ITER1})) :
\beq\label{ITER11}
W_i^{(0)}\,=\,\frac{v_i}{v\cdot P +i\gamma}\,,\eeq
which then leads to a polarization tensor which is neither
symmetric, nor transverse. For instance, the conductivity 
tensor $\sigma^{ij}$ built out of (\ref{ITER11}) reads:
\beq\label{SRTA}
\sigma_{ij}^{(0)}(\omega,{\bf p})\,=\,
im_D^2\int\frac{{\rm d}\Omega}{4\pi}\,\frac{v_i v_j}
{\omega - {\bf v\cdot p} + i\gamma}\,,\eeq
which is not transverse in the static limit:
%\footnote{Incidentally,
%a result like (\ref{SRTA}) has been recently reported
%in the literature \cite{Manuel99}, but this is merely due to some 
%incorrect algebraic manipulations in Ref. \cite{Manuel99}.}:
$p^i\sigma_{ij}^{(0)}(\omega=0,{\bf p})\ne 0$.

\subsection{Colour conductivities}

We now study the behaviour of the
polarization tensor $\Pi_{\mu\nu}(\omega, {\bf p})$ at very small
energy and momentum, $\omega,\,p\ll \gamma$. This is interesting since
the gradient expansion, which is only marginally justified for 
inhomogeneities at the scale $g^2 T$, becomes more and more accurate as 
the inhomogeneity becomes softer and softer. 
(Strictly speaking, colour inhomogeneities cannot 
be unambiguously defined at extremely soft scales $p\ll g^2T$,
for the reasons explained in the Introduction. The forthcoming discussion
is nevertheless interesting since it applies also for momenta
$p\sim g^2T$, at least within an expansion in powers of $1/\ln(1/g)$.)
By solving exactly the (linearized) Boltzmann equations (\ref{WLIN}) 
or (\ref{LIN}) in this kinematical limit,
we shall recover the previous result about Debye screening
(cf. eq.~(\ref{jstat})), and compute the longitudinal and transverse
colour conductivities defined in eq.~(\ref{SLT}). By ``exact solutions''
we mean here solutions which are obtained without
using the leading logarithmic approximation (\ref{LLPHI}) for
$\Phi({\bf v\cdot v}^\prime)$, and which are known 
up to corrections of O$(p/\gamma)$. Of course, these solutions account
only for the contributions of the hard and soft modes, with $q > \mu$,
to the corresponding conductivities. But they are still interesting
as they allow for the matching with the corresponding contributions
of the ultrasoft modes, to be computed non-perturbatively
(see eq.~(\ref{sigNLL}) below, and the discussion after it).

We shall study
the two following situations: ({\it i}) $\omega \ll p \ll \gamma$
(this includes the static case $\omega =0$ as a particular limit), and
({\it ii}) $p\ll \omega \simle \gamma$.
Since, generally, the electric and magnetic sectors 
behave differently in these limits, it is useful to project the Boltzmann
equation (\ref{LIN}) for $W^i(P,{\bf v})$
onto longitudinal and tranverse components (cf. eq.~(\ref{WL})):
\beq\label{WLT}
(v\cdot P)W_L(P,{\bf v})&=&{{\bf v} \cdot \hat{\bf p}}
\,-\,i\gamma\Bigl\{W_L(P,{\bf v})\,-\,\langle W_L(P,{\bf v})\rangle
\Bigr\}\nonumber\\
(v\cdot P)W_T^i(P,{\bf v})&=&v_T^i\,-\,
i\gamma\Bigl\{W^i_T(P,{\bf v})\,-\,\langle W_T^i(P,{\bf v})\rangle
\Bigr\}.\eeq

({\it i}) Consider first the {\it static} limit $\omega \to 0$, where
the longitudinal sector should provide Debye screening, as shown in Sec.
3.3. And, indeed, for $\omega =0$, the above equation for $W_L$ reduces to:
\beq\label{eqL0}
({{\bf v} \cdot \hat{\bf p}})(1+pW_L)\,=\,
 i\gamma\Bigl\{W_L(p,{\bf v})\,-\,\langle W_L(p,{\bf v})\rangle
\Bigr\},\eeq
with the obvious solution\footnote{This solution
can be obtained also from the iteration (\ref{ITER2}),
when written for $W_L$ and $\omega=0$: then, the zeroth order term 
reads $W_L^{(0)}(\omega=0)=-1/p$, which, being independent of
${\bf v}$, makes all the higher order corrections to vanish:
$W_L^{(N)}(\omega=0)$ for $N\ge 1$ (cf. the discussion prior to
eq.~(\ref{NULL})).} $W_L(\omega=0,p,{\bf v})=-1/p$.
Note that the collision term in eq.~(\ref{eqL0}) vanish identically
for this solution, which is therefore independent of $\gamma$
(and thus the same as in the HTL approximation).
Since, moreover, $W(\omega=0)=
p W_L(\omega=0) A_0({\bf p})$ for static fields (cf. eq.~(\ref{WI})),
this solution is clearly equivalent to $W(\omega=0,{\bf p},{\bf v})=
- A_0({\bf p})$, as expected from the discussion in Sec. 3.3.
When inserted into eq.~(\ref{PILT}), it yields:
\beq\label{PIL0}
\Pi_L(\omega=0,p)&=&m_D^2,\eeq 
which, together with the first equation (\ref{SLT}),
gives the behaviour of the longitudinal conductivity
$\sigma_L$ at small $\omega$ (and for arbitrary $p$):
\beq\label{SL0}
\sigma_L(\omega\to 0,p)\,\to\,-i\omega\,\frac{m_D^2}{p^2}\,.\eeq
The results (\ref{PIL0}) and (\ref{SL0}) are the same as in the
HTL approximation \cite{BIO96}: at small frequencies,
the electric sector is not affected by the collision effects
(see also Ref. \cite{ASY99}).

In the same limit, however, important modifications occur in the magnetic 
sector. Consider, indeed, eq.~(\ref{WLT}) for $W_T$ at $\omega =0$:
\beq\label{eqT0}
({{\bf v} \cdot {\bf p}})W_T^i\,=\,-\,v_T^i\,+\,
i\gamma\Bigl\{W^i_T({\bf p},{\bf v})\,-\,\langle W_T^i({\bf p, v})\rangle
\Bigr\}.\eeq
In the absence of the collision term (that is, to zeroth order in the
iteration (\ref{ITER2})), 
this equation would imply $W_T^{i\,(0)}=-v_T^i/({\bf v\cdot
p} - i\epsilon)$ (where the $i\epsilon$ stays for retarded boundary
conditions, as in eq.(\ref{GR})), which would then generate 
the small frequency limit of the HTL magnetic polarization
tensor \cite{BIO96,MLB96} :
\beq\label{HTL0}
\Pi_T^{(0)}(\omega\ll p)\,\simeq\,-i\,\frac{\pi}{2}\,\omega
m_D^2\int\frac{{\rm d}\Omega}{4\pi}\,v^i(\delta_{ij}-\hat p_i\hat p_j)
v^j\,\delta({\bf v\cdot p})\,=\,-i\,\frac{\pi}{2}\,\frac{\omega}{p}
\,m_D^2\,,\eeq
thus yielding $\sigma_T^{(0)}(\omega\ll p)\simeq (\pi/2)(m_D^2/p)$.
These expressions, which are formally singular as $p\to 0$, are
correct only as long as $p\gg \gamma$.
For ultrasoft momenta $p\simle \gamma$, they are modified by
the collision terms, as we discuss now.

Note first that, unlike in the electric sector where
$W_L(\omega=0)=-1/p$ is independent of ${\bf v}$, here $W_T^i({\bf p, v})$
is a non-trivial function of ${\bf v}$ already in the zeroth order
iteration.
For such a function, the collision term in the r.h.s. of eq.~(\ref{eqT0})
cannot vanish; it is thus a quantity of order $\gamma$, with 
respect to which the drift term in the l.h.s.
of (\ref{eqT0}) can be neglected in the longwavelength limit
$p \ll \gamma$. In this limit, the equation for $W_T$ reduces to:
\beq\label{T0}
v_T^i\,=\,
i\gamma\Bigl\{W^i_T({\bf v})\,-\,\langle W_T^i({\bf v})\rangle
\Bigr\}.\eeq
This is a simple equation which can be solved exactly. Specifically,
since $v_T^i$ is the only vector left in the problem, we can write:
$W^i_T({\bf v})= (C/i\gamma)v_T^i$ with some coefficient $C$.
Then,
\beq
\langle W_T^i({\bf v})\rangle \,=\, (C/i\gamma)\langle v_T^i \rangle
\,=\,\kappa(C/i\gamma)v_T^i,\eeq
where we have denoted $\kappa v^i\equiv \langle v^i\rangle$, so that
(cf. eq.~(\ref{AVF})) :
\beq\label{kappa}
\kappa\,=\,\frac{
\int\frac{{\rm d}\Omega'}{4\pi}\,\Phi({\bf v\cdot v}^\prime)\,
({\bf v\cdot v}')}{\int\frac{{\rm d}\Omega'}{4\pi}\,
\Phi({\bf v\cdot v}^\prime)}\,.\eeq
 Physically, $\kappa = \langle
\cos\alpha\rangle$, where $\alpha$ is the angle
made by the velocities  ${\bf v}$ and ${\bf v}'$ of the
colliding particles, and the brackets denote averaging with 
respect to the scattering cross section, cf. eq.~(\ref{AVF});
obviously, $|\kappa| < 1$.
Then, eq.~(\ref{T0}) fixes the coefficient $C$ as $C=1/(1-\kappa)$.
To conclude:
\beq\label{WT0} W_T^i({\bf v})\,=\,-\,\frac{i}{\gamma}\,
\frac{v_T^i}{1-\kappa}\,=\,-i\,\frac{v_T^i}{\gamma - \delta},\eeq
with (recall eq.~(\ref{gamma1})):
\beq\label{delta}
\delta\,\equiv\,\gamma\kappa\,=\,m_D^2\,\frac{g^2 N T}{2}
\int\frac{{\rm d}\Omega'}{4\pi}\,\Phi({\bf v\cdot v}^\prime)\,
({\bf v\cdot v}').\eeq
With the leading logarithmic approximation (\ref{LLPHI}) 
for $\Phi({\bf v\cdot v}^\prime)$ in eq.~(\ref{delta}), the angular
integral over ${\bf v}'$ vanishes by parity. Thus,
$\delta$ is a finite quantity of O$(g^2 T)$, which is completely
determined by the present approximation. Its explicit evaluation, however,
requires the full expression (\ref{PHII}) for $\Phi({\bf v\cdot v}^\prime)$.
We write $\delta \equiv  \alpha NT \bar\delta$, and obtain  $\bar\delta$
by numerical  integration of 
eq.~(\ref{delta}). The result is $\bar\delta=-0.20305024...\,$.

With $ W_T^i({\bf v})$ from eq.~(\ref{WT0}), it is finally straightforward
to estimate the transverse conductivity, or polarization tensor,
 in the present kinematical limit
(cf. eqs.~(\ref{PILT}) and (\ref{SLT})) :
\beq\label{ST0}
\sigma_T(\omega=0, p\to 0)&=&
\frac{i}{2}\,m_D^2\int\frac{{\rm d}\Omega}{4\pi}\,{\bf v\cdot W}_T({\bf v})
\,\simeq\,\frac{m_D^2}{3(\gamma - \delta)}
\,=\,\frac{\omega_{pl}^2}{\gamma - \delta}\,,\nonumber\\
\Pi_T(\omega\ll p \ll g^2 T)&\simeq&\,-i\omega_{pl}^2\,
\frac{\omega}{\gamma - \delta}\,,\eeq
where $\omega_{pl}\equiv m_D/\sqrt 3$ is
the {\it plasma frequency}, that is, the frequency of the longwavelength
($p\to 0$) collective excitations \cite{BIO96,MLB96}.
Strictly speaking, eqs.~(\ref{ST0}) hold only for 
very low momenta $p\ll g^2 T$; indeed, in their derivation
above, we have
neglected the drift term in the l.h.s. of (\ref{eqT0}), but we have
kept the term $\delta\sim g^2 T$ coming from the collision integral. 
In the limit where $\ln(1/g)$ is large, one can further expand these
expressions in powers of $\delta/\gamma \sim 1/\ln(1/g)$ and get, 
to linear order,
\beq\label{STEXP}
\sigma_T(\omega=0)\,\simeq\,
\frac{\omega_{pl}^2}{\gamma}\,\left(1 + \frac{\delta}{\gamma}\right),\eeq
where the neglected terms are down by, at least, two inverse powers
of $\ln(1/g)$. Remarkably, it turns out that, within the same accuracy, 
eq.~(\ref{STEXP}) holds also for momenta $p\sim g^2T$, which is
a case of physical interest\footnote{We thank Larry Yaffe for this remark.}.
Indeed, since $p/\gamma \sim 1/\ln(1/g)$ as well, one can solve
eq.~(\ref{eqT0}) by iterations, as a formal expansion in powers
of $\ln^{-1}\equiv 1/\ln(1/g)$. This yields (compare to eq.~(\ref{WT0}))
\beq\label{WTp} W_T^i(p, {\bf v})\,=\,-i\,\frac{v_T^i}{\gamma}\left
[1 +\frac{\delta}{\gamma}-i\frac{{\bf v\cdot p}}{\gamma}+
{\rm O}\left(\ln^{-2}\right)\right],\eeq
which shows that $p\sim g^2 T$ 
and $\delta$ enter on the same footing in $W_T^i$. But when constructing the
colour conductivity, as in eqs.~(\ref{PILT}) and (\ref{SLT}), the 
term  in eq.~(\ref{WTp}) which involves ${\bf v\cdot p}$ 
vanishes after angular integration, so we are left with the
same expression for $\sigma_T$ as above, eq.~(\ref{STEXP}).
By using eq.~(\ref{gammamu}) for $\gamma$, together with the numerical
estimate for $\delta$ given above, we can rewrite eq.~(\ref{STEXP}) 
in the following form:
\beq\label{sigNLL}
\sigma_T^{-1}(\omega\ll p\sim g^2T)\,=\,
\frac{\alpha NT}{\omega_{pl}^2}\left[\ln\frac{m_D}{\mu} -\bar\delta
+ {\rm O}\left(\ln^{-2}\right)\right].\eeq
This expression, which represents the contribution of the
hard and soft modes (with momenta $k\simge \mu$) to the colour
conductivity at the scale $g^2 T$, turns out to be useful for
the matching with the corresponding contribution of the 
ultrasoft modes \cite{AY}.

Eq.~(\ref{sigNLL}), which
holds to leading and next-to-leading order in an
expansion in powers of $\ln^{-1}$,
extends the solution to the Boltzmann equation
for $\sigma_T(\omega\ll p\sim g^2T)$ beyond the LLA of Sec. 3.4.
Note that, even to this order, the transverse colour conductivity
remains local (i.e., independent of the momentum $p$), as in the LLA.
But this is only true within the accuracy indicated in
eq.~(\ref{sigNLL}): the first corrections to this equation, 
of ${\rm O}(\ln^{-2})$, involve $p^2/\gamma^2$ and thus are non-local.

The expressions in eq.~(\ref{ST0}) should be also compared to 
the HTL result in eq.~(\ref{HTL0}): this shows that it is essentially 
the damping rate $\gamma$ which cuts off the divergence of $\Pi_T$, or
$\sigma_T$, as $p\to 0$. This is in line with Drude's picture
of the electric conductivity, and it is interesting
to pursue this comparison even further, so as to emphasize the
difference between colour and electric conductivity
(see also Ref. \cite{ASY98} for a discussion on this point).
The colour fluctuation induced by a uniform, and transverse,
colour mean field ${\bf E}^a_T$, as determined by the solution
above (cf. eqs.~(\ref{WT0}) and (\ref{WI})) :
\beq\label{WT01} 
W(p\ll \gamma, {\bf v})
\,\simeq\,\frac{{\bf v\cdot E}_T}{\gamma}\,\equiv\,
\tau_{col}\,{\bf v\cdot E}_T,\qquad
{\rm with}\,\,\,\,\tau_{col}\sim\,\frac{1}{\gamma}\,\sim\,
\frac{1}{g^2T\ln(1/g)}\,,
\eeq
should be
compared with the charge fluctuation induced by an electric field,
in the relaxation time approximation \cite{Baym97} (below,
$\alpha_{em}=e^2/4\pi$, with $e$ the electric charge):
\beq\label{WT02} 
W_{el}({\bf p})\,=\,\tau_{el}(p){\bf v\cdot E}_T,
\qquad {\rm with}\,\,\,\,\tau_{el}(p)\,\sim\,\frac{p}
{\alpha_{em}\ln(1/\alpha_{em}) T^2}\,\sim\,\frac{1}{
e^4T\ln(1/e)}\,.\eeq
Besides the loss of two powers of the coupling constant in the 
denominator
($\tau_{col} \sim 1/g^2$, as compared to $\tau_{el}\sim 1/e^4$),
the colour relaxation time appears to be independent of the
momentum of the hard particles (unlike $\tau_{el}$, which is
proportional to $p$). Thus, the present approximation for 
colour transport is formally similar to the relaxation time
approximation for electric charge, but with a
shorter, and momentum independent, relaxation time\footnote{We
thank Henning Heiselberg for a clarifying discussion on 
this point.}.
We emphasize, however, that in the case of colour this is
a property of the exact solution of the corresponding Boltzmann
equation, and not a consequence of the ``relaxation time
approximation''.

({\it ii}) Another interesting limit where the (linearized) Boltzmann
equation (\ref{LIN}) can be solved exactly is the longwavelength,
or zero momentum, limit, $p\to 0$, at fixed frequency $\omega$
(with $\omega \simle \gamma$, for the present approximations
to apply). Once again, an exact solution can be found because,
once $p$ is neglected, the vector fluctuation
$W^i({\bf v})$ is necessarily proportional to $v^i$.
Moreover, if the momentum $p$ is strictly zero, one cannot 
distinguish between longitudinal and transverse polarizations,
so that $W^i(\omega,p=0,{\bf v})=C(\omega) v^i$ must be the 
solution to (cf. eq.~(\ref{LIN})) :
\beq\label{LIN0}
\omega W^i(\omega,{\bf v})&=&v^i\,-\,
i\gamma\Bigl\{W^i(\omega,{\bf v})\,-\,\langle W^i(\omega,{\bf v})\rangle
\Bigr\}.\eeq
This is similar to the previous eq.~(\ref{T0}),
so that the corresponding solution reads (compare to eq.~(\ref{WT0})):
\beq W^i(\omega,p=0,{\bf v})\,=\,
\frac{v^i}{\omega+i(\gamma - \delta)},\eeq
which yields the following, isotropic, conductivity tensor:
\beq\label{SIGIJ}
\sigma^{ij}(\omega,p=0)\,=\,i\delta^{ij}\,
\frac{\omega_{pl}^2}{\omega+i(\gamma - \delta)}\,
\equiv\,\delta^{ij}\,\sigma(\omega,p=0).\eeq
In particular, as $\omega \ll \gamma$, we obtain
\beq
\sigma(p=0,\omega\ll\gamma)\,\simeq\,\frac{\omega_{pl}^2}
{\gamma - \delta}\,,\eeq
which is formally the same result as in the static case
(cf. eq.~(\ref{ST0})) except that it applies now to
both the longitudinal, and the transverse, conductivities:
$\sigma_L(p=0)=\sigma_T(p=0)=\sigma(p=0)$.
This should be compared with the previous results in 
eqs.~(\ref{SL0}) and (\ref{ST0}): 
unlike the longitudinal conductivity, the transverse one
appears to be continuous
in the double limit $\omega\to 0$ and $p\to 0$ (in the sense that
the two limits give identical results).

\section{Conclusions}

In this paper, we have used the Boltzmann equation describing
the relaxation of colour fluctuations in order to generate a set
of gauge-invariant amplitudes for the ultrasoft fields, i.e.,
the fields with momenta $\simle g^2T$. These amplitudes determine
the response of the hard quasiparticles to longwavelength
($\lambda \ge 1/g^2T$) colour mean fields. They define 
an effective theory for the ultrasoft fields, resulting from
integrating out the modes with momenta larger
than $g^2  T$ in perturbation theory.

The strategy which has been used in this paper,
namely the use of the kinetic theory in order to construct
effective amplitudes for the soft fields, is
reminiscent of our previous construction of the hard thermal
loops from collisionless kinetic equations \cite{qcd}.
The new element here is the inclusion of the effects of
the collisions, which is essential since these are leading order
effects for the colour excitations at the scale $g^2 T$.
This results in two important differences with respect to the
previous analysis of the HTL's: {\it a}) At a technical level,
kinetic theory appears to be the only workable approach toward
the systematic construction of the ultrasoft amplitudes. Indeed, 
unlike the HTL's, which are one-loop amplitudes and have been 
originally obtained in a diagrammatic approach \cite{BP90,FT90}, 
the ultrasoft amplitudes correspond to an infinite series of 
Feynman graphs which are conveniently resummed, to the order of 
interest, by the solution of the Boltzmann equation. {\it b})
As for their physical content, the main new ingredient
in the ultrasoft amplitudes is the effect of dissipation
as a consequence of collisions. 
In the leading logarithmic approximation, and also in the 
next-to-leading order (NLO) approximation as defined in eq.~(\ref{sigNLL}),
the dissipation is simply encoded in a local colour conductivity. 
In general, this is described by the (non-local) imaginary parts
of the ultrasoft amplitudes (as, e.g., in eq.~(\ref{PI1})).

By studying the Boltzmann equation, we have been able to obtain
a few exact results about the ultrasoft amplitudes, in particular,
the Ward identities they satisfy (cf. eq.~(\ref{Wtr})),
the static limit of the induced current (cf. eq.~(\ref{jstat})),
and the colour conductivity beyond the 
leading logarithmic approximation (cf. eqs.~(\ref{ST0}), 
(\ref{sigNLL})
and (\ref{SIGIJ})). More generally, by using formal solutions
obtained by iterations, we have studied the non-local structure
of the ultrasoft amplitudes (cf. eq.~(\ref{PI1})), and 
established their diagrammatic interpretation (cf. Sec. 3.2).

As already emphasized, the dynamics of the ultrasoft colour fields
described by the Boltzmann equation is dissipative: any initial
colour excitation will die away after a typical time
$\tau_{col}\sim 1/(g^2T\ln(1/g))$.
(This should be contrasted with the collisionless
dynamics in the HTL approximation, which is conservative \cite{qcd},
and even Hamiltonian \cite{Nair,baryo}.)
This dissipative description is appropriate to study the
relaxation of given initial off-equilibrium deviations in the average
colour density, as, e.g., in the calculation of the colour
conductivity, in Sec. 3.2. 

For many other applications --- for instance,
in studies of the baryon number violation in the hot electroweak
plasma ---, one is interested in colour excitations at the 
scale $g^2 T$ which are generated by thermal fluctuations 
in the plasma. The 
most convenient strategy to deal with such non-perturbative fluctuations is to treat them
as classical fields at finite temperature, which can then be
simulated on a lattice
\cite{McLerran,ASY96,Hu,baryo,Bodeker98,Moore98}.
In such a framework, one has to be able to also generate
the correct thermal correlations of the ultrasoft fields,
at least in the classical approximation.
In practice, and especially for numerical simulations, it is 
convenient to use a Langevin description
of the fluctuations, that is, to
simulate the thermal correlations with an 
appropriate ``noise'' term. The ``noise'' is a random source with zero 
expectation value but non-trivial correlators which are chosen
so as to induce, via the equations of motion, the proper
thermal correlations of the ultrasoft fields.

For the effective theory at the scale $g^2 T$,
the appropriate noise term has been constructed
by B\"odeker \cite{Bodeker98}. 
This term does not appear naturally in the derivation
of the Boltzmann equation from quantum field theory \cite{BE},
where one focuses on the distribution function of the hard
particles, rather than on the dynamics of the soft fields.
Still, to the order of interest, the structure of the noise
can be reconstructed from the known structure of the collision
term, by using the fluctuation dissipation theorem.
This construction will be presented somewhere else \cite{XI}.

{\bf Acknowledgements} We thank 
Fran\c coise Gu\'erin, Henning Heiselberg, Guy Moore and  Larry Yaffe
for useful discussions and comments on the manuscript. Many of these discussions took place during our stay at 
the Institute for Nuclear Theory, Washington University, which we thank for hospitality and support. Finally we are
grateful to  Tony Rebhan for his help with the numerical evaluation of $\bar\delta$.

%\newpage

\end{document}